\documentclass[prd,a4paper,nofootinbib,final,twocolumn,floatfix]{revtex4}
\usepackage{amsfonts}
\usepackage{amsmath}
\usepackage{amssymb}
\usepackage[usenames,dvips]{color}
\usepackage[english]{babel}
\usepackage[latin1]{inputenc}
\usepackage{amsfonts}
\usepackage{appendix}
\usepackage{epsfig}
\usepackage{graphics,rotating}
\usepackage{dcolumn}
\usepackage{bm}
\usepackage{color}
\usepackage[usenames,dvipsnames,svgnames]{xcolor}
\usepackage[colorlinks=true,
            linkcolor=blue,
            urlcolor=black,
            citecolor=blue]{hyperref}
\usepackage{hyperref}
\usepackage[T1]{fontenc}
\usepackage{multirow}
\usepackage{float}
\usepackage{subfigure}
\usepackage{enumitem}

\begin{document}

\title{Stability analysis of the spin evolution fix points in inspiraling
compact binaries with black hole, neutron star, gravastar, or boson star
components}
\author{Zolt\'an Keresztes$^{\dag }$, L\'aszl\'o \'Arp\'ad Gergely$^{\star }$%
}
\affiliation{Institute of Physics, University of Szeged, D\'om t\'er 9, Szeged 6720,
Hungary\\
$^{\dag }${\small E-mail: zkeresztes@titan.physx.u-szeged.hu\quad \quad }$%
^{\star }${\small \ E-mail: gergely@physx.u-szeged.hu }}

\begin{abstract}
Based on a recently derived secular spin evolution of black holes, neutron
stars, gravastars, or boson stars in precessing compact binaries on
eccentric orbit, we carry out a linear stability analysis of fix point
configurations. We identify the aligned and more generic coplanar
configurations of the spins and orbital angular momentum as fix points.
Through a dynamical system analysis, we investigate their linear stability
as function of the mass quadrupole parameter. Our most important results are
as follows. Marginal stability holds for the binary configurations with both
spins antialigned to the orbital angular momentum, for both spins aligned to
the orbital angular momentum (with the exception of certain quadrupolar
parameter ranges of neutron stars and boson stars), and for the extremal
mass ratio. For equal masses, the configurations of one of the spins aligned
and the other antialigned is stable for gravastar binaries, for neutron star
binaries in the high quadrupolar parameter range, and for boson star
binaries. For some unequal mass gravastar binaries, black hole binaries or
neutron star binaries, a transition from stability to instability can occur
during the inspiral, when one of the spins is aligned, while the other is
antialigned to the orbital angular momentum. We also discover a transitional
instability regime during the inspiral of certain gravastar, neutron star,
or boson star binaries with opposing spins. For coplanar configurations we
recover the marginally stable configurations leading to the libration
phenomenon identified in previous numerical investigations lacking mass
quadrupole contributions, and we analyze how it is affected by the
quadrupolar structure of the sources. We also investigate the linear
stability of black hole, neutron star, and boson star binaries, also of
mixed black hole - gravastar, black hole - neutron star and black hole -
boson star binaries. We find instabilities only for the gravastar -
gravastar, boson star - boson star and black hole - boson star binaries. For
a given spin configuration, marginal stability strongly depends on the value
of the quadrupolar parameters. The stability region is larger for neutron
star binaries than for black hole binaries, while the mixed systems have a
restricted stability parameter region.
\end{abstract}

\maketitle

\section{Introduction}

Compact binaries are among the most probable sources of gravitational
radiation. All gravitational wave detections by the Advanced LIGO and Virgo
detectors up to now have been from these type of sources, mostly from black
hole - black hole systems \cite{LIGOcatO12}, \cite{GraceDB}, but also
neutron star - neutron star \cite{LIGO5NS} with multimessenger counterparts
and black hole - lighter compact object (which is either the lightest black
hole or the heaviest neutron star yet discovered) \cite{LIGOunequal}.

Compact binary evolution can be well described analytically when the
separation of the components is large as compared to their Schwarzschild
radius. During this inspiral regime, the Keplerian evolution is modified by
general relativistic corrections at first and higher post-Newtonian (PN)
orders \cite{damourderuelle}, \cite{blanchet}. The spin of the compact
objects also contributes at 1.5 PN orders through the spin-orbit (SO) and at
2PN orders through the spin-spin (SS) couplings \cite%
{barkeroconnell2,barkeroconnell1,barkeroconnell,KWW,KIDDER52,RS,GPV3,GSS,GSS2,KJ,BFH}
. The rotation of the compact object also induces a mass deformation,
characterized by its quadrupole. The coupling of this to the mass of the
other component enters at 2PN orders in the dynamics \cite%
{poisson,GKQM,MVG,Racine}. The 2PN order accurate instantaneous dynamics has
been discussed in detail in Refs. \cite{Inspiral1,Inspiral2,chameleon}.
Finally gravitational radiation kicks in at 2.5 PN orders \cite{Peters}.

In this paper, we continue the study of the secular spin dynamics in compact
binaries on eccentric orbit during the inspiral regime on the conservative
timescale, thus to 2PN order accuracy, commenced in our earlier work \cite%
{KTG}, which will be referred as Paper I. The binary components are compact
objects with masses $m_{i}$, dimensionless spins $\chi _{i}$, and
quadrupolar parameter $w_{i}$ spanning\ over a wide range of values, being $%
1 $ by definition for black holes, falling into $\left( -0.8,1\right) $ for
gravastars \cite{GravaStarw}, into $\left( 2,14\right) $ for neutron stars 
\cite{NSw}, \cite{NSUrbanecw}, and into $\left( 10,150\right) $ for boson
stars \cite{BosonStarw}. In Paper I, we have derived a closed system of
first-order differential equations for the secular evolution of the spin
polar angles $\kappa _{i}$ (taken in the system with the orbital angular
momentum $\mathbf{L}_{\mathbf{N}}$ on the $z$ axis) and the difference of
their azimuthal angles $\Delta \zeta $ (measured in the orbital plane).%
\footnote{%
If we were to use an alternative set of angular variables defined with the
direction of the total orbital angular momentum $\mathbf{L}$, rather than
the Newtonian angular momentum $\mathbf{L}_{\mathbf{N}}$, differences would
be induced only through the SO contribution to L (as the PN and 2PN
contributions lie in the direction of $\mathbf{L}_{\mathbf{N}}$). For a
discussion in terms of angles related to $\mathbf{L}$ of the inspiral
waveforms for the coplanar resonant configurations \cite{Schnittman} and
consequences in data analysis, see Ref. \cite{Gupta}.} We have analyzed both
analytically and numerically various particular situations with special
emphasis on the effect of the quadrupole parameter (thus on the role of the
companion of a black hole) manifesting itself in the flip-flopping evolution
of the polar angles. In this paper, we will concentrate on the stability
analysis of fixed points of the spin angle dynamics, again emphasizing the
role played by the quadrupole parameter.

The early analysis \cite{KIDDER52} of the orbit-averaged (secular)
spin-precession equations for circular orbits, with only the SO and SS
contributions included, identified four collinear spin and angular momentum
equilibrium configurations (up-up, up-down, down-up, and down-down).
Coplanar configurations representing equilibrium solutions of the
instantaneous angular evolutions were also found \cite{Schnittman}. There, a
numerical study of the evolution of the phase shift (the difference in the
azimuthal angles of the spins) revealed solutions librating about the
equilibrium (interpreted as stable) and unstable solutions departing from
the equilibrium. The latter appeared as a sequence of "long periods of
stasis followed by short bursts of rapid divergence away from equilibrium",
dubbed as quasistable. Both these analyses disregarded the mass quadrupole -
mass monopole (QM) contributions, driven by the parameters $w_{i}$.

When the rotation-induced quadrupole of black holes is also taken into
account ($w_{i}=1$), the four above-mentioned collinear configurations still
represent equilibrium \cite{Racine}. Among these, the up-down configuration
was found unstable in certain parameter regimes \cite{GKSKBDT}.

In this paper, we revisit the analysis of the fixed points and their
stability in the more generic context of compact binaries with black hole,
neutron star, gravastar, or boson star components, thus allowing for an
arbitrary value of the quadrupole parameters $w_{i}$.

In Sec. \ref{closed} we sum up the closed system of evolution equations of
the spin angles and we establish the notations.

In Sec. \ref{collinear} we discuss this system from the dynamical systems
point of view. We prove that the configurations of spins aligned and
antialigned to the orbital angular momentum are still fixed points of the
evolution in this more generic setup. We analyze their linear stability by
deriving the marginal stability conditions. Then we also discuss the equal
and unequal mass configurations. For the former, we identify new types of
instabilities, while for the latter, we identify a new type of evolution
during inspiral, encompassing a transitional instability.

We proceed with the analysis of the fixed points in Sec. \ref{coplanar},
investigating coplanar configurations, discussing the dependence on the
quadrupolar parameters of the marginally stable configurations. Further, we
investigate here the linear stability of black hole, neutron star, and boson
star binaries, also of mixed black hole - gravastar, black hole - neutron
star, and black hole - boson star binaries.

In Sec. \ref{concludingr} we present the conclusions.

\section{Secular spin angle evolutions\label{closed}}

The secular evolution of the spin angles $\kappa _{1}$, $\kappa _{2}$ and $%
\Delta \zeta \equiv \zeta _{1}-\zeta _{2}$ has been derived in Paper I in
the form of a closed first-order differential system:\footnote{%
For notational simplicity, we omit the overbar (present in Paper I) from the
secular time derivatives. We follow this simplified notation in what
follows, keeping in mind that the equations describe secular dynamics.}%
\begin{eqnarray}
\frac{1}{R}\frac{d\kappa _{1}}{d\mathfrak{t}} &=&\left( 1+\nu -x_{1}\cos
\kappa _{1}-\nu w_{2}x_{2}\cos \kappa _{2}\right)  \notag \\
&&\times x_{2}\sin \kappa _{2}\sin \Delta \zeta ~,  \label{eq1}
\end{eqnarray}%
\begin{eqnarray}
\frac{1}{R}\frac{d\kappa _{2}}{d\mathfrak{t}} &=&-\left( 1+\nu
^{-1}-x_{2}\cos \kappa _{2}-\nu ^{-1}w_{1}x_{1}\cos \kappa _{1}\right) 
\notag \\
&&\times x_{1}\sin \kappa _{1}\sin \Delta \zeta ~,  \label{eq2}
\end{eqnarray}%
\begin{eqnarray}
\frac{1}{R}\frac{d\Delta \zeta }{d\mathfrak{t}} &=&\nu -\nu ^{-1}+\left(
1+2\nu ^{-1}-w_{1}\right.  \notag \\
&&\left. -w_{1}\nu ^{-1}x_{1}\cos \kappa _{1}\right) x_{1}\cos \kappa _{1} 
\notag \\
&&-\left( 1+2\nu -w_{2}-w_{2}\nu x_{2}\cos \kappa _{2}\right) x_{2}\cos
\kappa _{2}  \notag \\
&&-\left( 1+\nu ^{-1}-w_{1}\nu ^{-1}x_{1}\cos \kappa _{1}\right)  \notag \\
&&\times x_{1}\cot \kappa _{2}\sin \kappa _{1}\cos \Delta \zeta  \notag \\
&&+\left( 1+\nu -w_{2}\nu x_{2}\cos \kappa _{2}\right)  \notag \\
&&\times x_{2}\cot \kappa _{1}\sin \kappa _{2}\cos \Delta \zeta  \notag \\
&&-x_{1}x_{2}\left( \frac{\sin \kappa _{2}}{\sin \kappa _{1}}-\frac{\sin
\kappa _{1}}{\sin \kappa _{2}}\right) \cos \Delta \zeta ~,  \label{eq3}
\end{eqnarray}%
with the notations%
\begin{equation}
R=\frac{3\eta \pi }{\mathfrak{T}~\mathfrak{\bar{l}}_{r}^{2}}~,~x_{i}=\frac{%
\chi _{i}}{\mathfrak{\bar{l}}_{r}}~.
\end{equation}%
Here, $\nu =m_{2}/m_{1}\leq 1$ is the mass ratio, and $\eta =\mu /m$ is the
symmetric mass ratio (with $m=m_{1}+m_{2}$ and $\mu =m_{1}m_{2}/m$ the total
and reduced masses of the binary, respectively). Further, $\mathfrak{l}%
_{r}=cL_{N}/Gm\mu ~$is the dimensionless orbital angular momentum, and $%
\mathfrak{\bar{l}}_{r}$ is its average over a radial period ($G$, $c$, and $%
L_{N}$ denoting the gravitational constant, the speed of light, and the
magnitude of $\mathbf{L}_{\mathbf{N}}$, respectively). Finally, $\mathfrak{T}
$ is the radial period expressed to 2 PN accuracy as Eq. (6) of Paper I. In
addition the derivatives are with respect to a dimensionless time variable $%
\mathfrak{t}=tc^{3}/Gm$ (with time $t$) introduced in Ref. \cite{chameleon}.

Note that $1/\mathfrak{l}_{r}^{2}$ represents one relative PN\ order, as
indicated by Eq. (8) of Ref. \cite{chameleon}. Hence in the angular
evolutions (\ref{eq1})-(\ref{eq3}), the $\mathcal{O}\left( R\right) $, $%
\mathcal{O}\left( Rx_{i}\right) $, and $\mathcal{O}\left( Rx_{i}x_{j}\right) 
$ terms are PN, 1.5PN, and 2PN contributions, respectively.

\section{Collinear spin orientations: linear stability analysis\label%
{collinear}}

The closed system (\ref{eq1})-(\ref{eq3}) becomes ill behaved for the
aligned configurations $\sin \kappa _{i}=0$. This is the usual singularity
of the polar angle in a spherical system of coordinates, for which the
azimuthal angles $\zeta _{i}$ become ill defined.

The case when only one of $\sin \kappa _{i}$ evolves through zero is
discussed in Appendix B of Paper I. We showed that despite the apparent
coordinate singularity, the system passes through such a configuration
driven by a well-defined dynamics.

When both $\sin \kappa _{1}$ and $\sin \kappa _{2}$ vanish the spins remain
parallel or antiparallel to the orbital angular momentum, i.e., the angles $%
\kappa _{i}$ being constant,\footnote{%
A similar conclusion valid only for black holes ($w_{i}=1$) also emerges
from Eqs. (3.2a) and (3.2b) of Ref. \cite{Racine}.} as can be seen from Eqs.
(\ref{eq1}) and (\ref{eq2}). With these the aligned configurations fulfill
the conditions for the fixed points of the spin dynamics. Remarkably, all
terms containing the problematic angles $\zeta _{i}$ in the evolution
equations of the Euler angles vanish; thus, their dynamics remains well
defined. We proceed with the linear stability analysis of these aligned
configurations.

We denote the fixed points as $\kappa _{\left( 0\right) i}$ and the points
obtained by a slight deviation $\epsilon _{i}\delta \kappa _{i}\left( 
\mathfrak{t}\right) $ as 
\begin{equation}
\kappa _{i}\left( \mathfrak{t}\right) =\kappa _{\left( 0\right) i}+\epsilon
_{i}\delta \kappa _{i}\left( \mathfrak{t}\right) ~,
\end{equation}%
with $\epsilon _{i}=\cos \kappa _{\left( 0\right) i}$ representing a\ sign
(this choice ensures that the perturbed angle stays in the domain $\left[
0,\pi \right] $). To leading order, the closed system becomes%
\begin{equation}
\frac{d\Delta \zeta }{d\mathfrak{t}}=A+\left( B_{1}\frac{\delta \kappa _{2}}{%
\delta \kappa _{1}}-B_{2}\frac{\delta \kappa _{1}}{\delta \kappa _{2}}%
\right) \cos \Delta \zeta ~,  \label{Dzeta_1}
\end{equation}%
\begin{equation}
\frac{d\delta \kappa _{1}}{d\mathfrak{t}}=B_{1}\delta \kappa _{2}\sin \Delta
\zeta ~,  \label{kappa1_1}
\end{equation}%
\begin{equation}
\frac{d\delta \kappa _{2}}{d\mathfrak{t}}=-B_{2}\delta \kappa _{1}\sin
\Delta \zeta ~,  \label{kappa2_1}
\end{equation}%
with the constants defined as%
\begin{eqnarray*}
\frac{A}{R} &=&\nu -\nu ^{-1} \\
&&+\epsilon _{1}\left( 1+2\nu ^{-1}-w_{1}-\epsilon _{1}\nu
^{-1}w_{1}x_{1}\right) x_{1} \\
&&-\epsilon _{2}\left( 1+2\nu -w_{2}-\epsilon _{2}\nu w_{2}x_{2}\right)
x_{2}~,
\end{eqnarray*}%
\begin{equation*}
\frac{B_{1}}{R}=\left[ \epsilon _{1}\left( 1+\nu -\epsilon _{2}\nu
w_{2}x_{2}\right) -x_{1}\right] x_{2}~,
\end{equation*}%
\begin{equation}
\frac{B_{2}}{R}=\left[ \epsilon _{2}\left( 1+\nu ^{-1}-\epsilon _{1}\nu
^{-1}w_{1}x_{1}\right) -x_{2}\right] x_{1}~.
\end{equation}

Time derivatives of Eqs. (\ref{Dzeta_1})--(\ref{kappa2_1}) combined lead to
the simple second-order differential equation 
\begin{equation}
\frac{d^{2}\mathcal{Z}}{d\mathfrak{t}^{2}}+\left( A^{2}+4B_{1}B_{2}\right) 
\mathcal{Z}=0
\end{equation}%
in the variable%
\begin{equation}
\mathcal{Z}=\delta \kappa _{1}\delta \kappa _{2}\sin \Delta \zeta ~,
\end{equation}%
with independent solutions $\mathcal{Z}_{\pm }=e^{\pm i\Omega \mathfrak{t}}$%
, where $\Omega ^{2}\equiv 4B_{1}B_{2}+A^{2}$. Obviously for $\Omega
^{2}\leq 0$ there is a runaway solution and the fixed point is unstable,
whereas for $\Omega ^{2}>0$, the independent solutions become harmonic
functions, which could be combined into 
\begin{equation}
\mathcal{Z}=Q\cos \left( \Omega \mathfrak{t}+G\right) ~,
\end{equation}%
with $Q$ and $G$ constants.

Then, Eq. (\ref{kappa1_1}) becomes 
\begin{eqnarray}
\frac{d\left( \delta \kappa _{1}\right) ^{2}}{d\mathfrak{t}} &=&2B_{1}\delta
\kappa _{1}\delta \kappa _{2}\sin \Delta \zeta  \notag \\
&=&2B_{1}Q\cos \left( \Omega \mathfrak{t}+G\right) ~,
\end{eqnarray}%
which results in%
\begin{equation}
\left( \delta \kappa _{1}\right) ^{2}=F_{1}+\frac{2B_{1}Q}{\Omega }\sin
\left( \Omega \mathfrak{t}+G\right) ~,  \label{kappa1sol_1}
\end{equation}%
with another integration constant $F_{1}\geq \left\vert 2B_{1}Q/\Omega
\right\vert $. Similarly, 
\begin{equation}
\left( \delta \kappa _{2}\right) ^{2}=F_{2}-\frac{2B_{2}Q}{\Omega }\sin
\left( \Omega \mathfrak{t}+G\right) ~,  \label{kappa2sol_2}
\end{equation}%
with the integration constant $F_{2}\geq \left\vert 2B_{2}Q/\Omega
\right\vert $. The solutions (\ref{kappa1sol_1}) and (\ref{kappa2sol_2})
prove that the aligned configurations $\sin \kappa _{1,2}=0$ are marginally
stable for $\Omega ^{2}>0$.

\subsection{Sufficient conditions for marginal stability\label{suff}}

The above stability criterion depends on seven parameters:%
\begin{equation}
\nu ,~x_{i},~w_{i},~\epsilon _{i}~.
\end{equation}%
We discuss their possible ranges below. The mass ratio is $\nu \in (0,1]$.
The value\ of $\mathfrak{\bar{l}}_{r}$ can be estimated by Eq. (8) of Ref. 
\cite{chameleon}, which for not too eccentric bound orbits is of the order $%
\mathfrak{\bar{l}}_{r}\geq 1/\sqrt{\bar{\varepsilon}}$ (equality holds for
circular orbits \cite{chameleon}). Assuming $\chi _{i}\leq 1$ and that the
PN approximation is valid for the PN parameter $\bar{\varepsilon}\in \left(
0,0.1\right) $ gives for circular orbits $x_{i}=\chi _{i}/\mathfrak{\bar{l}}%
_{r}\in \left( 0,0.3\right) $.

First, we note that the positivity of $B_{i}$ implies%
\begin{equation}
\epsilon _{1}\epsilon _{2}w_{3-i}<\epsilon _{1}\epsilon _{2}\frac{1+\nu
^{2i-3}\left( 1-\epsilon _{i}x_{i}\right) }{\epsilon _{3-i}x_{3-i}}~.
\end{equation}%
The signs of $B_{1}$ and $B_{2}$ are shown in Table \ref{table1}. 
\begin{table}[tbp]
\begin{center}
\begin{tabular}{|c||c|c|}
\hline\hline
& $B_{1}>0$ & $B_{2}>0$ \\ \hline\hline
$%
\begin{array}{c}
\kappa _{\left( 0\right) 1}=0 \\ 
\kappa _{\left( 0\right) 2}=0%
\end{array}%
$ & $w_{2}<W_{2}^{-}$ & $w_{1}<W_{1}^{-}$ \\ \hline
$%
\begin{array}{c}
\kappa _{\left( 0\right) 1}=0 \\ 
\kappa _{\left( 0\right) 2}=\pi%
\end{array}%
$ & Always & $w_{1}>W_{1}^{+}$ \\ \hline
$%
\begin{array}{c}
\kappa _{\left( 0\right) 1}=\pi \\ 
\kappa _{\left( 0\right) 2}=0%
\end{array}%
$ & $w_{2}>W_{2}^{+}$ & Always \\ \hline
$%
\begin{array}{c}
\kappa _{\left( 0\right) 1}=\pi \\ 
\kappa _{\left( 0\right) 2}=\pi%
\end{array}%
$ & Never & Never \\ \hline
\end{tabular}%
\end{center}
\caption{The positivity conditions for $B_{1}$ and $B_{2}$. A sufficient
condition for marginal stability is met when their signs agree.}
\label{table1}
\end{table}
There, we introduced the notations%
\begin{equation}
W_{i}^{\pm }=\frac{1+\nu ^{3-2i}\left( 1\pm x_{3-i}\right) }{x_{i}},
\label{Wi}
\end{equation}%
which for the allowed parameter ranges are bound as%
\begin{eqnarray}
W_{1}^{-} &>&\frac{10}{3}~,\quad W_{2}^{-}>\frac{17}{3}~,~  \notag \\
W_{1}^{+} &>&\frac{10}{3}~,\quad W_{2}^{+}>\frac{20}{3}~.  \label{MSa}
\end{eqnarray}%
Note that as $\mathfrak{\bar{l}}_{r}^{-1}$ increases monotonically during
the inspiral the derivatives 
\begin{equation}
\frac{dW_{i}^{\pm }}{d\left( \mathfrak{\bar{l}}_{r}^{-1}\right) }=-\frac{%
1+\nu ^{3-2i}}{\chi _{i}}\mathfrak{\bar{l}}_{r}^{2}<0~  \label{Wilr}
\end{equation}%
ensure that the functions $W_{i}^{\pm }$ decrease monotonically.

For extreme mass ratios $\nu \ll 1$, we immediately obtain $\left\vert
B_{1}B_{2}\right\vert \ll A^{2}$, resulting in $\Omega ^{2}>0$, and thus the
fixed point of the system is marginally stable. In addition, for $%
B_{1}B_{2}>0$, the system is also marginally stable.

Finally, there is marginal stability also in the case when at least one the $%
B_{i}$ vanishes, unless $A=0$ also holds. When both spins are nonvanishing,
this occurs for the pair of quadrupolar parameters%
\begin{equation}
w_{i}=\frac{1+\nu ^{3-2i}\left( 1-\epsilon _{3-i}x_{3-i}\right) }{\epsilon
_{i}x_{i}}\in \left\{ \pm W_{i}^{\pm }\right\} ~,
\end{equation}%
among which only $w_{i}\in \left\{ W_{i}^{\pm }\right\} \,$\ are admitted on
physical grounds. Note that the vanishing of $B_{i}$ and $A$ also implies
the vanishing of $B_{3-i}$. Hence, in order to have marginal stability, both 
$B_{i}$ cannot vanish simultaneously.

If only one spin is present, 
\begin{eqnarray}
x_{3-i} &=&0  \notag \\
w_{i} &=&\frac{\nu ^{3-2i}-\nu ^{2i-3}+\epsilon _{i}\left( 1+2\nu
^{2i-3}\right) x_{i}}{\left( \epsilon _{i}+\nu ^{2i-3}x_{i}\right) x_{i}}~.
\label{wi}
\end{eqnarray}%
Note that in this case $\Omega ^{2}=A^{2}$ is a convex function, everywhere
positive outside its double degenerate root $w_{i}$, which represents its
minimum.

For equal masses, Eq. (\ref{wi}) reduces to 
\begin{eqnarray}
x_{3-i} &=&0  \notag \\
w_{i} &=&\frac{3}{1+\epsilon _{i}x_{i}}~.  \label{wicr2}
\end{eqnarray}

There are the following cases to be discussed:

\begin{enumerate}
\item[i)] When both spins are antialigned with the orbital angular momentum,
the system is marginally stable. All other aligned configurations require
additional conditions to be imposed on the parameters.

\item[ii)] In the configuration of the compact binary components with both
spins aligned, $w_{1}<10/3$ together with~$w_{2}<17/3$ implies $%
w_{i}<W_{i}^{-}$ during the whole inspiral; hence, marginal stability holds
for all mass ratios and spin values. Therefore, the aligned black hole -
black hole, gravastar - gravastar, and black hole - gravastar binary systems
are marginally stable.

\item[iii)] For neutron star binaries with both spins aligned, the marginal
stability depends on their equation of state. For low $w_{i}$ values, both $%
B_{i}\geq 0$ and $B_{3-i}>0$ could hold true throughout the inspiral, in
this case the system being marginally stable. Similarly, marginal stability
could hold throughout the inspiral for larger values of $w$ of either
aligned neutron star binaries or aligned boson star binaries, when both $%
B_{i}\leq 0$ and $B_{3-i}<0$ hold. There could be cases when marginal
stability during the initial phases of the inspiral turns into instability
and possibly turns back to stability again, or initial instability turns
into marginal stability, as $W_{i}^{-}$ evolve during the inspiral.

\item[iv)] With one spin aligned and another antialigned, the marginal
stability criterion is $w_{i}>W_{i}^{+}$ for the aligned spin. This could
hold either during the whole inspiral or only at its latter stages, when the
decreasing $W_{i}^{+}$ may slide below a high enough value of the
quadrupolar parameter $w_{i}$, even if it does not hold in the earlier
stages of the inspiral.
\end{enumerate}

\subsection{Equal masses}

For equal masses%
\begin{eqnarray}
\frac{\Omega ^{2}}{R^{2}} &=&\sum_{i=1}^{2}\left[ \left( 1+x_{i}^{2}\right)
w_{i}^{2}-6w_{i}+9\right.  \notag \\
&&\left. +2\epsilon _{i}w_{i}\left( w_{i}-3\right) x_{i}\right]
x_{i}^{2}+2\left( 2+w_{1}w_{2}\right) x_{1}^{2}x_{2}^{2}  \notag \\
&&-\sum_{i=1\left( j\neq i\right) }^{2}2\epsilon _{i}\left(
4+w_{j}+w_{i}w_{j}\right) x_{i}x_{j}^{2}  \notag \\
&&-2\epsilon _{1}\epsilon _{2}x_{1}x_{2}\left[ 1+w_{1}w_{2}-\sum_{i=1}^{2}%
\left( 3+2x_{i}^{2}\right) w_{i}\right] ~.  \label{Omsq}
\end{eqnarray}

\subsubsection{Equal spins and quadrupolar parameters}

Equation (\ref{Omsq}) reduces for $w_{1}=w_{2}=w$ and $x_{1}=x_{2}=x=\chi /%
\mathfrak{\bar{l}}_{r}$ to%
\begin{eqnarray}
\frac{\Omega ^{2}}{R^{2}} &=&2\left[ \left( 3-w\right) ^{2}+2\left(
1+w^{2}\right) x^{2}\right] x^{2}  \notag \\
&&-2\epsilon _{1}\epsilon _{2}\left( 1-6w+w^{2}-4wx^{2}\right) x^{2}  \notag
\\
&&-8\left( \epsilon _{1}+\epsilon _{2}\right) \left( 1+w\right) x^{3}~.
\label{Om2}
\end{eqnarray}

For $\epsilon _{1}=\epsilon _{2}=\epsilon $, Eq. (\ref{Om2}) reduces to 
\begin{equation}
\frac{\Omega ^{2}}{R^{2}}=4x^{2}\left( x+wx-2\epsilon \right) ^{2}~,
\label{Om3}
\end{equation}%
and $\Omega ^{2}$ vanishes for%
\begin{equation}
w=w_{cr_{1}}=\frac{2\epsilon }{x}-1~,  \label{wcr1}
\end{equation}%
but apart from this critical value, it is always positive, and thus the
configuration is marginally stable. Then:

\begin{enumerate}
\item[a)] For $\epsilon =-1$, $w_{cr_{1}}<-1$, which is outside of the
astrophysically interesting range. The configuration with both spins
antialigned to the orbital angular momentum is marginally stable.

\item[b)] For $\epsilon =1$, 
\begin{equation}
\frac{dw_{cr_{1}}}{d\left( \mathfrak{\bar{l}}_{r}^{-1}\right) }=-\frac{2}{%
\chi }\mathfrak{\bar{l}}_{r}^{2}<0~,
\end{equation}%
shows that $w_{cr1}$ decreases monotonically during the inspiral, attaining
its minimal value at the end of it. The lowest value occurs at the end of
the inspiral on circular orbits, leading to the bound $w_{cr_{1}}>17/3$.
Thus the critical value may fall into the possible range of the quadrupole
parameter for neutron stars and boson stars. Apart from this value, the
configuration with both spins aligned to the orbital angular momentum is
marginally stable.

\item[c)] For $\epsilon _{1}\epsilon _{2}=-1$,%
\begin{equation}
\frac{\Omega ^{2}}{R^{2}}=\allowbreak 4x^{2}\left( w-1\right) \left(
w+wx^{2}-x^{2}-5\right) ~
\end{equation}%
and $\Omega ^{2}$ has two roots:%
\begin{equation}
w_{-}=1~,~w_{+}=1+\frac{4}{x^{2}+1}\in \left( 4.7,5\right) ~.
\end{equation}%
The angular frequency $\Omega ^{2}>0$ for $w>w_{+}$ and $w<1$, and $\Omega
^{2}<0$ for $1<w<w_{+}$. Hence, for the configurations with one spin aligned
and another antialigned to the orbital angular momentum, gravastar binaries
are marginally stable, while black hole binaries are unstable. For neutron
star binaries, these configurations can be either unstable or marginally
stable (depending on the equation of state), while for boson star binaries,
they are always marginally stable. These findings are compatible with and
expand those described under iv) in Sec. \ref{suff} for the case of equal
masses, spins, and quadrupole parameters.
\end{enumerate}

\subsubsection{Second spin negligible}

In another limiting case, $x_{2}\ll x_{1}$, the expression (\ref{Omsq})
reduces to%
\begin{equation}
\frac{\Omega ^{2}}{R^{2}}=x_{1}^{2}\left( 3-w_{1}-\epsilon
_{1}w_{1}x_{1}\right) ^{2}~,
\end{equation}%
which is positive, thus yielding marginal stability, except for%
\begin{equation}
w_{1}=w_{cr_{2}}^{\epsilon _{1}}=\frac{3}{1+\epsilon _{1}x_{1}}~,
\end{equation}%
which reproduces the earlier result (\ref{wicr2}). Thus, there is a critical
value of $w_{1}$ as a function of $x_{1}$ where the configuration becomes
unstable. The critical value ranges as $w_{cr_{2}}\in \left( 2.3,3\right) $
for $\epsilon _{1}=1$ and as $w_{cr_{2}}\in \left( 3,4.3\right) $ for $%
\epsilon _{1}=-1$. Both can emerge only in a neutron star binary.

\begin{figure}[th]
\includegraphics[scale=0.65]{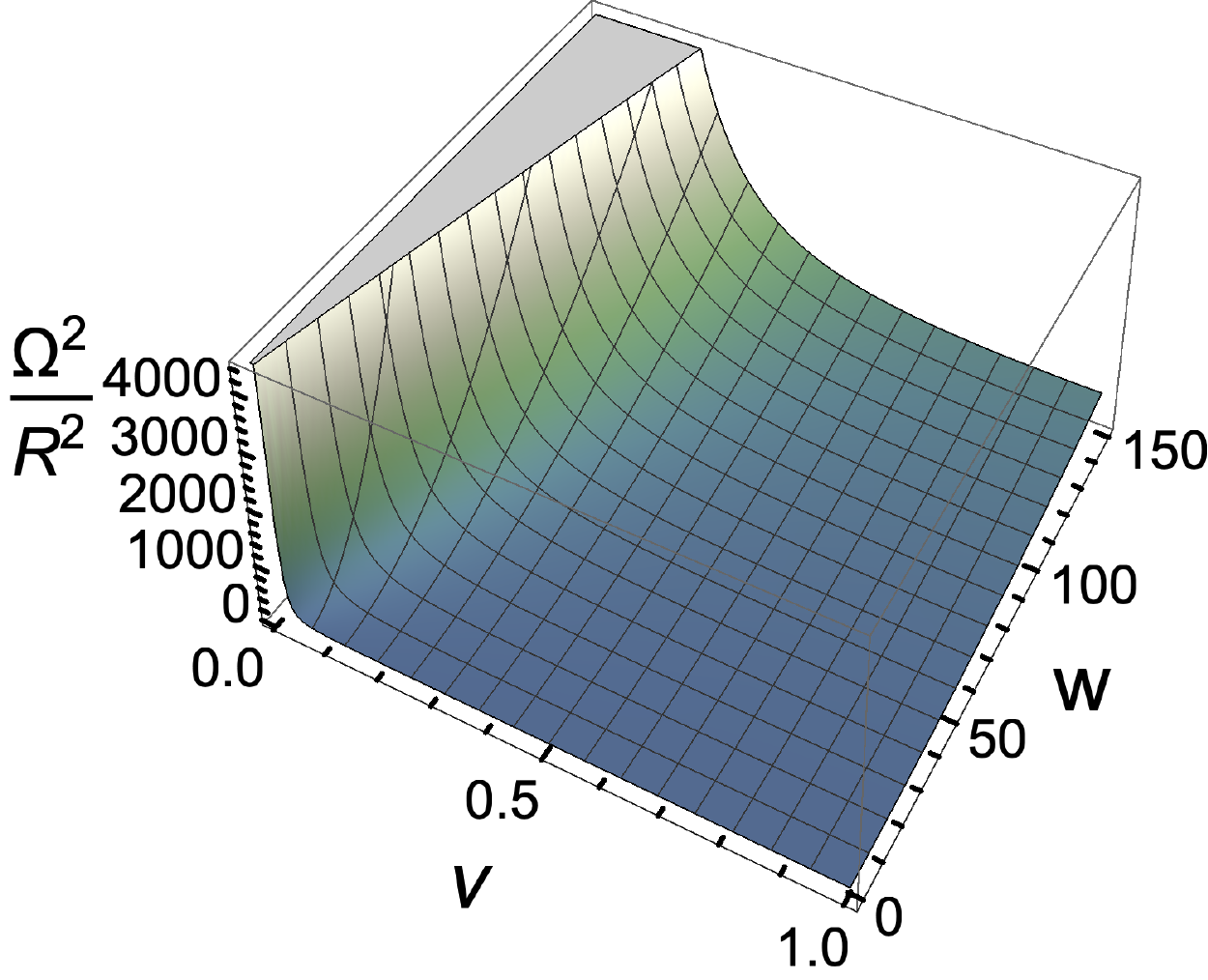} 
\centering
\caption{Marginal stability generically holds in the nonequal mass case with
both spins aligned to the orbital angular momentum, irrespective of the
values of $\protect\nu $, $w$ (shown) and $x$ (represented for $x=0.3$).}
\label{corotMS}
\end{figure}

\subsection{Nonequal masses in the nonextreme mass ratio regime}

For simplicity, we set $w_{1}=w_{2}=w$ and $x_{1}=x_{2}=x$. Then, the
configurations with $\epsilon _{1}=\epsilon _{2}$ are marginally stable.
Indeed, when both spins are counterrotating, $\epsilon _{1}=\epsilon _{2}=-1$%
, and this has been already analytically proven in a more generic context
(see the item i) in Sec. \ref{suff}), while for the corotating case, $%
\epsilon _{1}=\epsilon _{2}=1$, the stability condition reads 
\begin{gather}
\frac{\Omega ^{2}}{R^{2}}=4\left( 2+\nu +\nu ^{-1}\right) \left( 1-x\right)
\left( 1-wx\right) x^{2}  \notag \\
+4\!\left( 1\!-\!w\right) \!^{\!2}\!x^{4}+\left( \nu \!-\!\nu ^{-1}\right)
\!^{\!2}\!\left( 1\!-\!2x\!+\!wx^{2}\right) \!^{\!2}>0~.
\end{gather}%
There are no real roots of the equation $\Omega ^{2}=0$, and numerical
analysis has proven that these configurations are marginally stable. We
illustrate this for $x=0.3$ on Fig. \ref{corotMS}.

\begin{figure*}[th]
\includegraphics[scale=0.5]{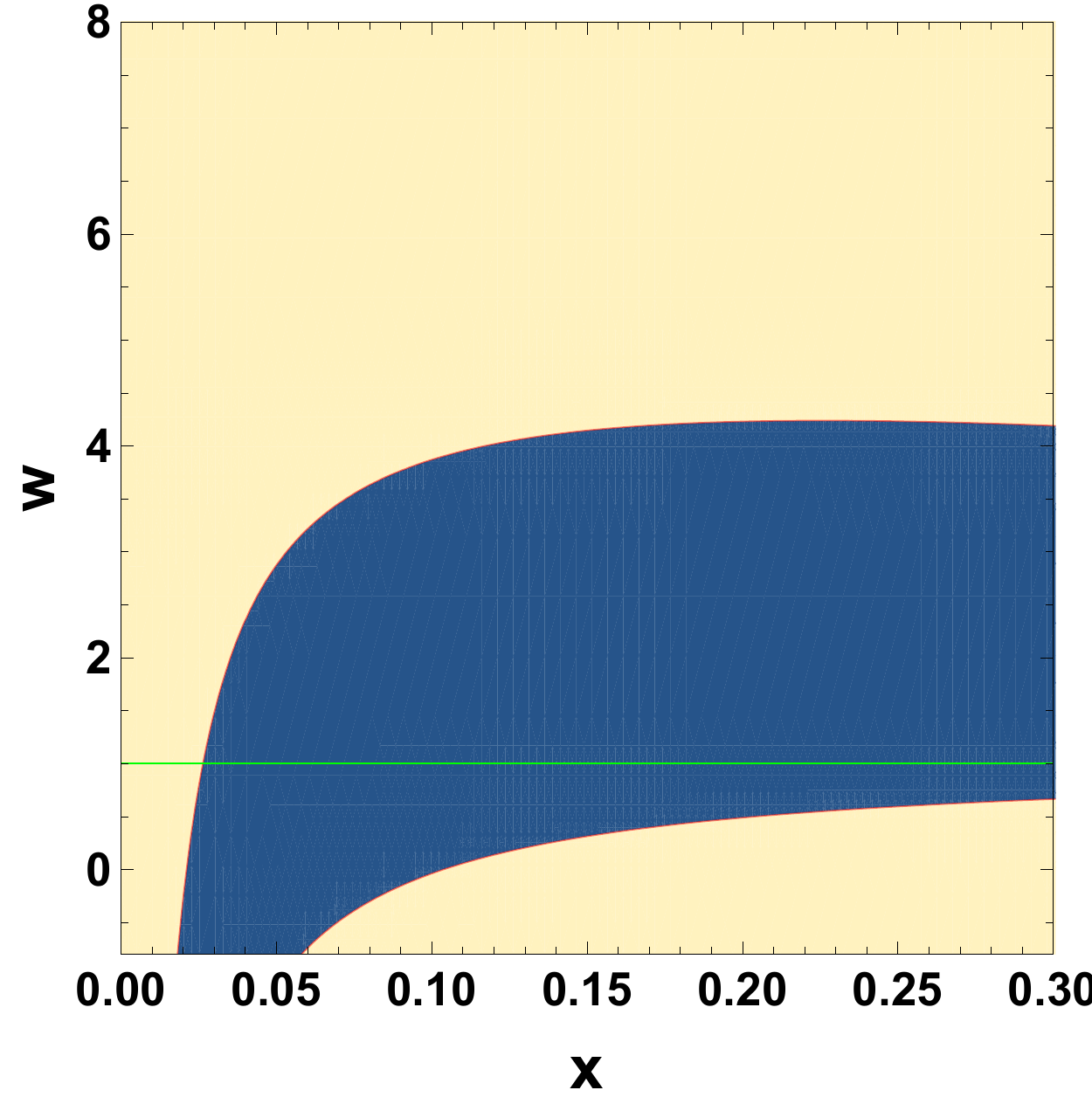} %
\includegraphics[scale=0.510]{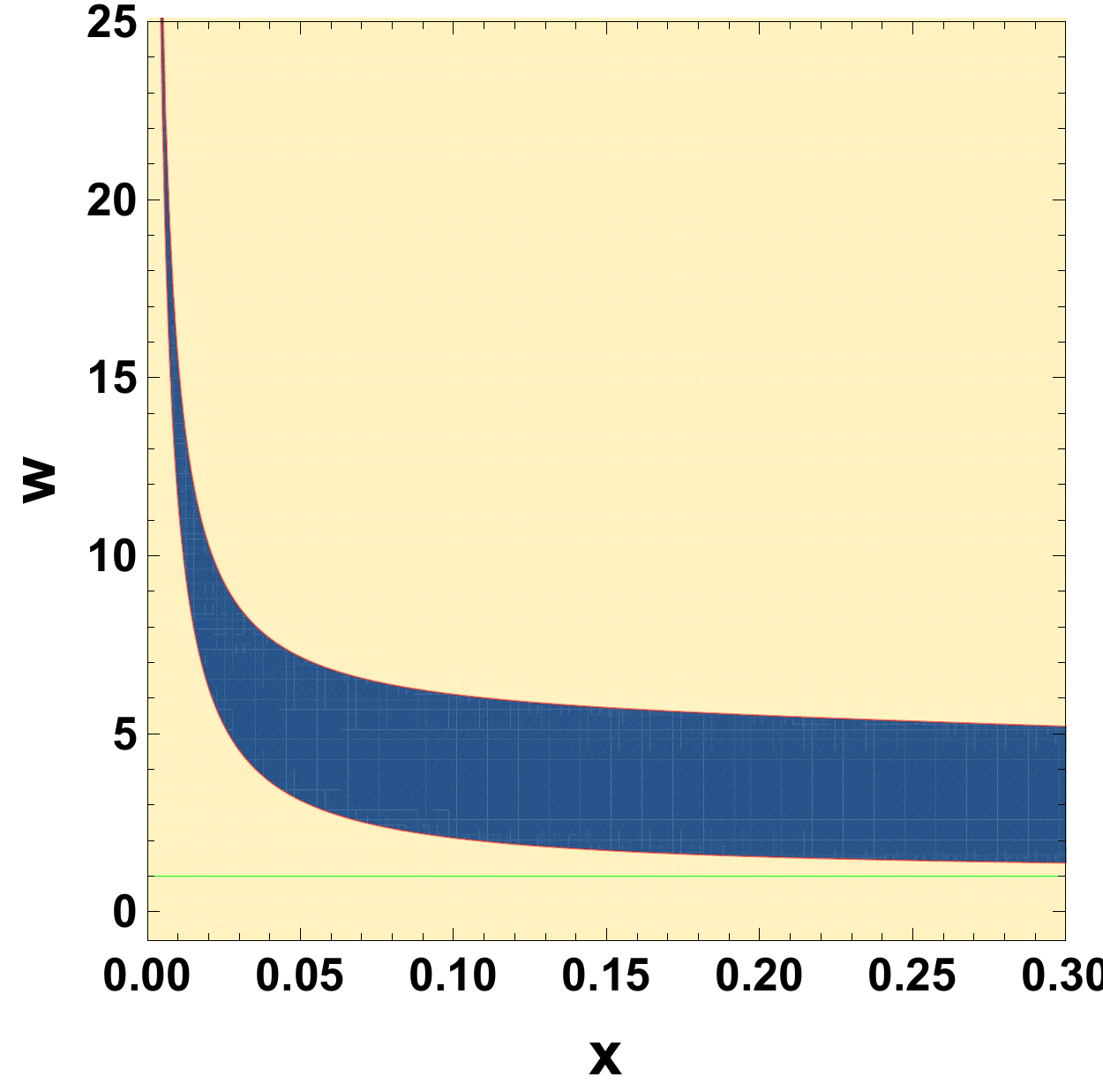} \centering
\caption{Marginal stability regions for the quadrupolar parameter $w$
represented by positive $\Omega ^{2}$ (shown in yellow) as opposed to
instability regions (blue) for $\protect\epsilon _{1}=1$, $\protect\epsilon %
_{2}=-1$ (left panel) and $\protect\epsilon _{1}=-1$, $\protect\epsilon %
_{2}=1$ (right panel). The horizontal green lines are for black hole
binaries. Various transitions from stability to instability and back to
stability are possible as the inspiral proceeds (with increasing PN
parameter and $x$).}
\label{Contour1}
\end{figure*}

The cases $\epsilon _{1}=-\epsilon _{2}$ are shown on Fig. \ref{Contour1},
for the mass ratio $\nu =0.9$. The blue domain represents $\Omega ^{2}<0$,
thus unstable configurations, while in the yellow area, $\Omega ^{2}>0$. For
values of $w$ not shown on the figures [$w>8$ (left panel) and $w>25$ (right
panel)], $\Omega ^{2}>0$ holds (with the exception of the continuation of
the red curve), and hence marginal stability occurs.

The green horizontal lines on the panels represent black hole binaries. The
configuration with the larger mass black hole spin aligned to the orbital
angular momentum and the smaller mass black hole antialigned ($\epsilon
_{1}=1$, $\epsilon _{2}=-1$) is initially marginally stable, becoming
unstable as the inspiral orbit shrinks, as found in Ref. \cite{GKSKBDT}.

For neutron star binaries in the quadrupole parameter range $w\in \left(
2,4\right) $, linear stability can disappear abruptly during the inspiral
for any of the ($\epsilon _{1}=1$, $\epsilon _{2}=-1$) and ($\epsilon
_{1}=-1 $, $\epsilon _{2}=1$) configurations, similarly to the black hole
binaries.

By contrast, neutron star binaries with $w>4$ in the ($\epsilon _{1}=1$, $%
\epsilon _{2}=-1$) configuration are always marginally stable.

There are also configurations allowing for a sequence of evolutions, which
are stable, then unstable, then stable again during the inspiral. In these
cases, the instability is only transitional. Such systems are:

\begin{enumerate}
\item[A)] the gravastar binaries with $w$ values in the lower part of their
allowed range, in the configuration ($\epsilon _{1}=1$, $\epsilon _{2}=-1$);

\item[B)] neutron star binaries with $w$ values in the higher part of their
allowed range, in the ($\epsilon _{1}=-1$, $\epsilon _{2}=1$) configuration;

\item[C)] boson star binaries, also in the ($\epsilon _{1}=-1$, $\epsilon
_{2}=1$) configuration. For the latter, the instability occurs only for a
very limited part of the evolution (across the red curve in the right panel
of Fig. \ref{Contour1}).
\end{enumerate}

Finally note that the location and extension of the blue domain depends on
the mass ratio.

\begin{figure*}[th]
\includegraphics[scale=0.65]{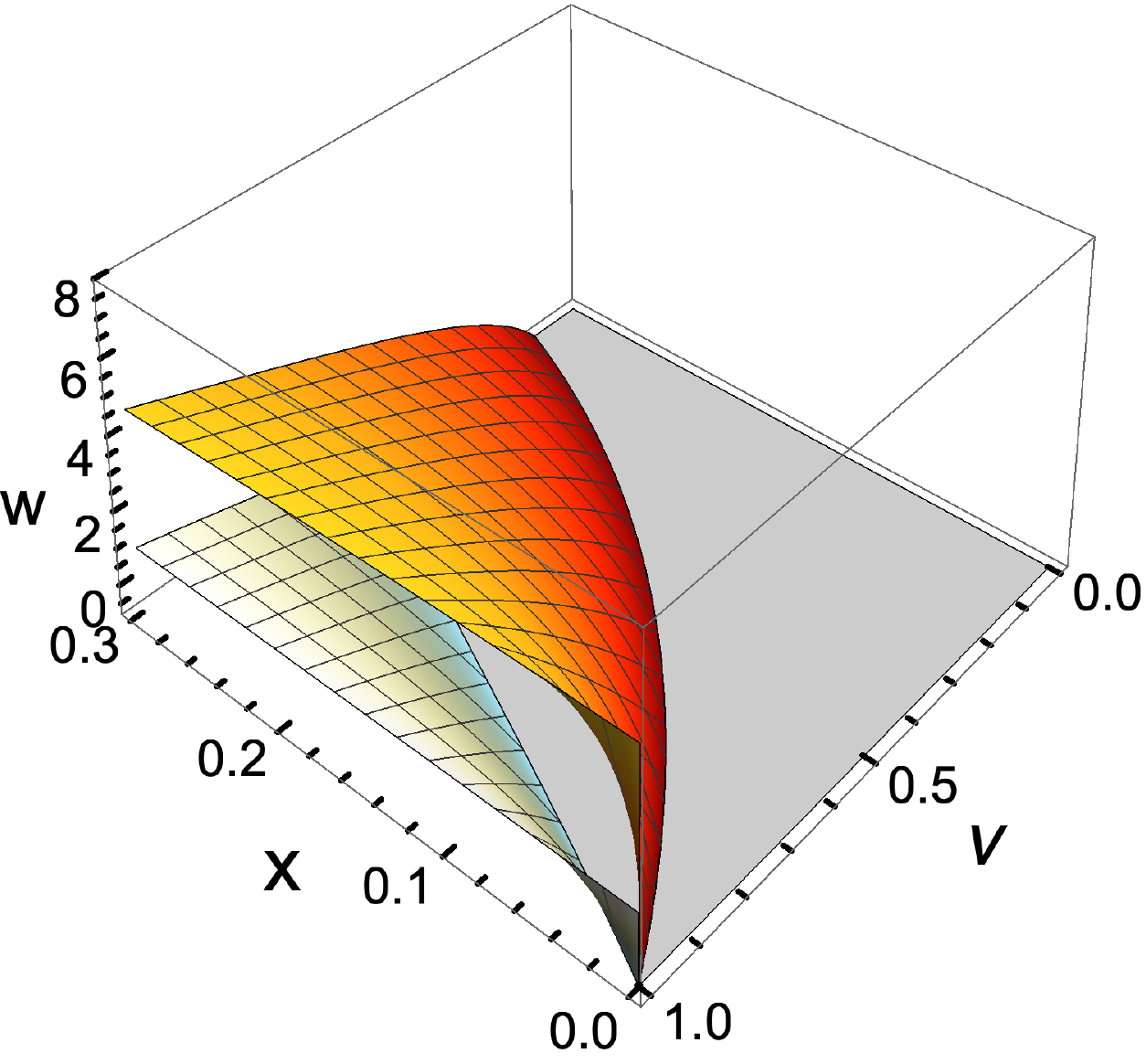} %
\includegraphics[scale=0.6]{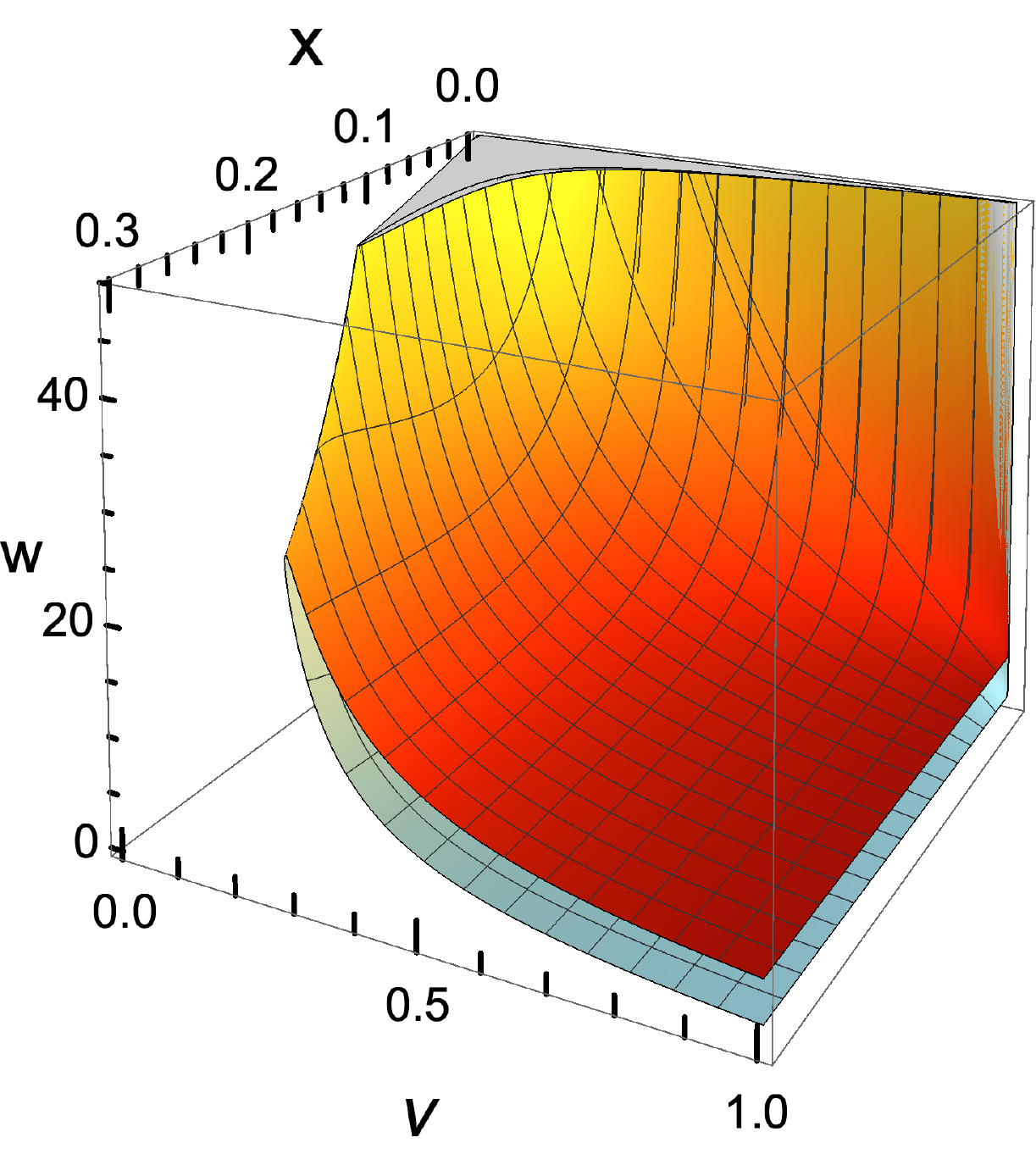} \centering
\caption{Marginal stability regions for the quadrupolar parameter $w$
represented by positive $\Omega ^{2}$ regions lying outside the two surfaces 
$\Omega ^{2}=0$, for $\protect\epsilon _{1}=1$, $\protect\epsilon _{2}=-1$
(left panel) and $\protect\epsilon _{1}=-1$, $\protect\epsilon _{2}=1$
(right panel). Marginal stability generically holds for low mass ratios $%
\protect\nu $, while for higher values of $\protect\nu $, regions of
stability and instability alternate, depending on the value of the
quadrupolar parameter $w$.}
\label{noneqmass}
\end{figure*}

We discuss the marginal stability for other mass ratio values below. For 
\begin{equation}
\epsilon _{2}=-\epsilon _{1}~,\quad w_{1}=w_{2}=w~,\quad x_{1}=x_{2}=x
\end{equation}%
stability is determined by the sign of 
\begin{eqnarray}
\frac{\Omega ^{2}}{R^{2}} &=&\allowbreak \left( \nu -\nu ^{-1}\right) ^{2}+%
\left[ 2-w\left( \nu +\nu ^{-1}\right) \right] ^{2}x^{4}  \notag \\
&&+4\epsilon _{1}\left( \nu -\nu ^{-1}\right) \left( 1+\nu +\nu
^{-1}-w\right) \left( 1+\allowbreak wx^{2}\right) x  \notag \\
&&-\allowbreak 4\epsilon _{1}\left( \nu -\nu ^{-1}\right) \left( 1+w\right)
x^{3}+2\left[ 2w^{2}\right.  \notag \\
&&+2\left( \nu ^{2}+\nu +1+\nu ^{-1}+\nu ^{-2}\right)  \notag \\
&&\left. +w\left( \nu ^{2}-4\nu -6-4\nu ^{-1}+\nu ^{-2}\right) \right]
x^{2}~.
\end{eqnarray}%
The roots of the $\Omega ^{2}=0$ equation are 
\begin{equation}
w_{\pm }^{\left( \epsilon _{1},\epsilon _{2}=-\epsilon _{1}\right) }=\frac{%
U_{1}+\epsilon _{1}U_{2}\pm \allowbreak V\sqrt{1+\epsilon _{1}x\left( \nu
^{-1}-\nu \right) }}{W_{1}+\epsilon _{1}W_{2}}
\end{equation}%
$\allowbreak $with$\allowbreak $%
\begin{eqnarray}
U_{1} &=&\left[ -1+4\nu +6\nu ^{2}+4\nu ^{3}-\nu ^{4}\right.  \notag \\
&&\left. +2x^{2}\nu \left( 1+\nu ^{2}\right) \right] x~,  \notag \\
U_{2} &=&-2\nu \left( 1-\nu ^{2}\right) +2\left( 1-\nu ^{4}\right) x^{2}~, 
\notag \\
V &=&\allowbreak 2\nu \left[ \left( 1+\nu \right) ^{2}-\epsilon _{1}\left(
1-\nu ^{2}\right) x\right] x~,  \notag \\
W_{1} &=&\left[ 4\nu ^{2}+\left( 1+\nu ^{2}\right) ^{2}x^{2}\right] x~, 
\notag \\
W_{2} &=&4\nu \left( 1-\nu ^{2}\right) x^{2}~.
\end{eqnarray}%
The two $\Omega ^{2}=0$ surfaces are represented on Fig. \ref{noneqmass}
(left panel for $\epsilon _{1}=1$, and right panel for $\epsilon _{1}=-1$).
In the region between the surfaces $\Omega ^{2}<0$, they are unstable
regions. For lower values of $\nu $, either there are no real roots, hence
no $\Omega ^{2}=0$ surfaces (for $\epsilon _{1}=-1$ this happens below $\nu
=\left( \sqrt{4x^{2}+1}-1\right) /2x$) or the roots are outside the
physically allowed region for the quadrupolar parameter (for $\epsilon
_{1}=1)$; thus, marginal stability holds irrespective of the value of $w$.
For larger values of $\nu $, a similar structure of the stability and
instability regions as for $\nu =0.9$ emerges, with the lower stability
region possibly missing for certain parameter combinations. 

\section{Coplanar spin orientation: fixed points and linear stability
analysis \label{coplanar}}

There are additional fixed points of the system (\ref{eq1})--(\ref{eq3})
given by%
\begin{equation}
\frac{d\kappa _{i}}{d\mathfrak{t}}=0~,~\frac{d\Delta \zeta }{d\mathfrak{t}}%
=0~.
\end{equation}%
The first condition is satisfied for $i=1$ and $i=2$, respectively, with%
\begin{eqnarray}
0 &=&\left( 1+\nu -x_{1}\cos \kappa _{1}-\nu w_{2}x_{2}\cos \kappa
_{2}\right)  \notag \\
&&\times x_{2}\sin \kappa _{2}\sin \Delta \zeta ~  \label{fix1}
\end{eqnarray}%
and 
\begin{eqnarray}
0 &=&\left( 1+\nu ^{-1}-x_{2}\cos \kappa _{2}-\nu ^{-1}w_{1}x_{1}\cos \kappa
_{1}\right)  \notag \\
&&\times x_{1}\sin \kappa _{1}\sin \Delta \zeta ~,  \label{fix2}
\end{eqnarray}%
while the second one is satisfied with%
\begin{eqnarray}
0 &=&\nu -\nu ^{-1}+\left( 1+2\nu ^{-1}-w_{1}\right.  \notag \\
&&\left. -\nu ^{-1}w_{1}x_{1}\cos \kappa _{1}\right) x_{1}\cos \kappa _{1} 
\notag \\
&&-\left( 1+2\nu -w_{2}-\nu w_{2}x_{2}\cos \kappa _{2}\right) x_{2}\cos
\kappa _{2}  \notag \\
&&-\left( 1+\nu ^{-1}-\nu ^{-1}w_{1}x_{1}\cos \kappa _{1}\right)  \notag \\
&&\times x_{1}\cot \kappa _{2}\sin \kappa _{1}\cos \Delta \zeta  \notag \\
&&+\left( 1+\nu -\nu w_{2}x_{2}\cos \kappa _{2}\right)  \notag \\
&&\times x_{2}\cot \kappa _{1}\sin \kappa _{2}\cos \Delta \zeta  \notag \\
&&-x_{1}x_{2}\left( \frac{\sin \kappa _{2}}{\sin \kappa _{1}}-\frac{\sin
\kappa _{1}}{\sin \kappa _{2}}\right) \cos \Delta \zeta ~.  \label{fix3}
\end{eqnarray}%
To investigate the stability about the fixed point ($\kappa _{\left(
0\right) 1}$, $\kappa _{\left( 0\right) 2}$, $\Delta \zeta _{\left( 0\right)
}$) under the perturbations of the angles, we parametrize them as%
\begin{equation}
\kappa _{1}\left( \mathfrak{t}\right) =\kappa _{\left( 0\right) 1}+\delta
\kappa _{1}\left( \mathfrak{t}\right) ~,
\end{equation}%
\begin{equation}
\kappa _{2}\left( \mathfrak{t}\right) =\kappa _{\left( 0\right) 2}+\delta
\kappa _{2}\left( \mathfrak{t}\right) ~,
\end{equation}%
\begin{equation}
\Delta \zeta \left( \mathfrak{t}\right) =\Delta \zeta _{\left( 0\right)
}+\delta \Delta \zeta \left( \mathfrak{t}\right) ~.
\end{equation}%
The perturbations $\left\vert \delta \kappa _{1}\right\vert $, $\left\vert
\delta \kappa _{2}\right\vert $, and $\left\vert \delta \Delta \zeta
\right\vert $ are initially much smaller than $1$, and if they stay so, the
configuration is stable or marginally stable. Equations (\ref{fix1}) and (%
\ref{fix2}) are solved by $\sin \kappa _{1,2}=0$ or $\sin \Delta \zeta =0$.
The first condition represents collinear spin configurations discussed
earlier, while the second yields the coplanar case. In this latter case, Eq.
(\ref{fix3}) reduces to the constraint 
\begin{eqnarray}
0 &=&\nu -\nu ^{-1}+\left( 1+2\nu ^{-1}-w_{1}\right.  \notag \\
&&\left. -\nu ^{-1}w_{1}x_{1}\cos \kappa _{\left( 0\right) 1}\right)
x_{1}\cos \kappa _{\left( 0\right) 1}  \notag \\
&&-\left( 1+2\nu -w_{2}-\nu w_{2}x_{2}\cos \kappa _{\left( 0\right)
2}\right) x_{2}\cos \kappa _{\left( 0\right) 2}  \notag \\
&&-\epsilon _{\Delta \zeta }\left( 1+\nu ^{-1}-\nu ^{-1}w_{1}x_{1}\cos
\kappa _{\left( 0\right) 1}\right)  \notag \\
&&\times x_{1}\cot \kappa _{\left( 0\right) 2}\sin \kappa _{\left( 0\right)
1}  \notag \\
&&+\epsilon _{\Delta \zeta }\left( 1+\nu -\nu w_{2}x_{2}\cos \kappa _{\left(
0\right) 2}\right) x_{2}\cot \kappa _{\left( 0\right) 1}\sin \kappa _{\left(
0\right) 2}  \notag \\
&&-\epsilon _{\Delta \zeta }x_{1}x_{2}\left( \frac{\sin \kappa _{\left(
0\right) 2}}{\sin \kappa _{\left( 0\right) 1}}-\frac{\sin \kappa _{\left(
0\right) 1}}{\sin \kappa _{\left( 0\right) 2}}\right) ~  \label{constr}
\end{eqnarray}%
between $\kappa _{\left( 0\right) 1}$ and $\kappa _{\left( 0\right) 2}$,
where $\epsilon _{\Delta \zeta }=\pm 1$ represents the sign of $\cos \Delta
\zeta _{\left( 0\right) }$. The simplest case of $\nu =1$, $w_{1}=w_{2}$,
and $x_{1}=x_{2}$ is solved for $\kappa _{\left( 0\right) 1}=\kappa _{\left(
0\right) 2}$.

To linear order in the perturbations about the fixed point, the evolution
equations (\ref{eq1}), (\ref{eq2}), and (\ref{eq3}) give%
\begin{equation}
\frac{d\delta \kappa _{1}}{d\mathfrak{t}}=A_{1}\delta \Delta \zeta ~,
\end{equation}%
\begin{equation}
\frac{d\delta \kappa _{2}}{d\mathfrak{t}}=-A_{2}\delta \Delta \zeta ~,
\end{equation}%
\begin{equation}
\frac{d\delta \Delta \zeta }{d\mathfrak{t}}=A_{3}\delta \kappa
_{1}-A_{4}\delta \kappa _{2}~,
\end{equation}%
with%
\begin{equation}
\frac{A_{1}}{R}=\left( 1+\nu -x_{1}\cos \kappa _{\left( 0\right) 1}-\nu
w_{2}x_{2}\cos \bar{\kappa}_{2}\right) x_{2}\sin \bar{\kappa}_{2}~,
\end{equation}%
\begin{eqnarray}
\frac{A_{3}}{R} &=&-\left( 1+2\nu ^{-1}-w_{1}\right.  \notag \\
&&\left. -2\nu ^{-1}w_{1}x_{1}\cos \kappa _{\left( 0\right) 1}\right)
x_{1}\sin \kappa _{\left( 0\right) 1}  \notag \\
&&-\epsilon _{\Delta \zeta }\left( 1+\nu ^{-1}\right) x_{1}\cot \kappa
_{\left( 0\right) 2}\cos \kappa _{\left( 0\right) 1}  \notag \\
&&+\epsilon _{\Delta \zeta }\nu ^{-1}w_{1}x_{1}^{2}\cot \kappa _{\left(
0\right) 2}\cos 2\kappa _{\left( 0\right) 1}  \notag \\
&&-\epsilon _{\Delta \zeta }\left( 1+\nu -\nu w_{2}x_{2}\cos \kappa _{\left(
0\right) 2}\right) x_{2}\frac{\sin \kappa _{\left( 0\right) 2}}{\sin
^{2}\kappa _{\left( 0\right) 1}}  \notag \\
&&+\epsilon _{\Delta \zeta }x_{1}x_{2}\frac{\cos \bar{\kappa}_{1}\left( \sin
^{2}\kappa _{\left( 0\right) 1}+\sin ^{2}\kappa _{\left( 0\right) 2}\right) 
}{\sin ^{2}\kappa _{\left( 0\right) 1}\sin \kappa _{\left( 0\right) 2}}~.
\end{eqnarray}%
The additional constants $A_{2}$ and $A_{4}$ are obtained from $A_{1}$ and $%
A_{3}$, respectively, with the changes $1\leftrightarrow 2$ and $\nu
\rightarrow \nu ^{-1}$. From this system, a second-order decoupled
differential equation for $\delta \Delta \zeta $ can be derived as 
\begin{equation}
\frac{d^{2}\delta \Delta \zeta }{d\mathfrak{t}^{2}}=\left(
A_{1}A_{3}+A_{2}A_{4}\right) \delta \Delta \zeta ~.
\end{equation}%
For $\omega ^{2}\equiv -\left( A_{2}A_{4}+A_{1}A_{3}\right) >0$, the
evolutions of $\delta \Delta \zeta $, $\delta \kappa _{1}$, and $\delta
\kappa _{2}$ are described by harmonic functions yielding marginally stable
fixed points. (In the terminology of Ref. \cite{Schnittman}, the
configuration is stable when $\Delta \zeta $ evolves through a harmonic
function about the equilibrium configuration.) Deviations from such
marginally stable configurations generate the librations.

Fix points with $\omega ^{2}\leq 0$ are unstable. The sign of $\omega ^{2}$
depends on the parameters%
\begin{equation}
\nu ,~x_{i},~w_{i}~,\kappa _{\left( 0\right) i}~,~\epsilon _{\Delta \zeta }
\label{params}
\end{equation}%
which are subject to the constraint (\ref{constr}).

\begin{figure*}[th]
\begin{center}
\includegraphics[scale=0.5]{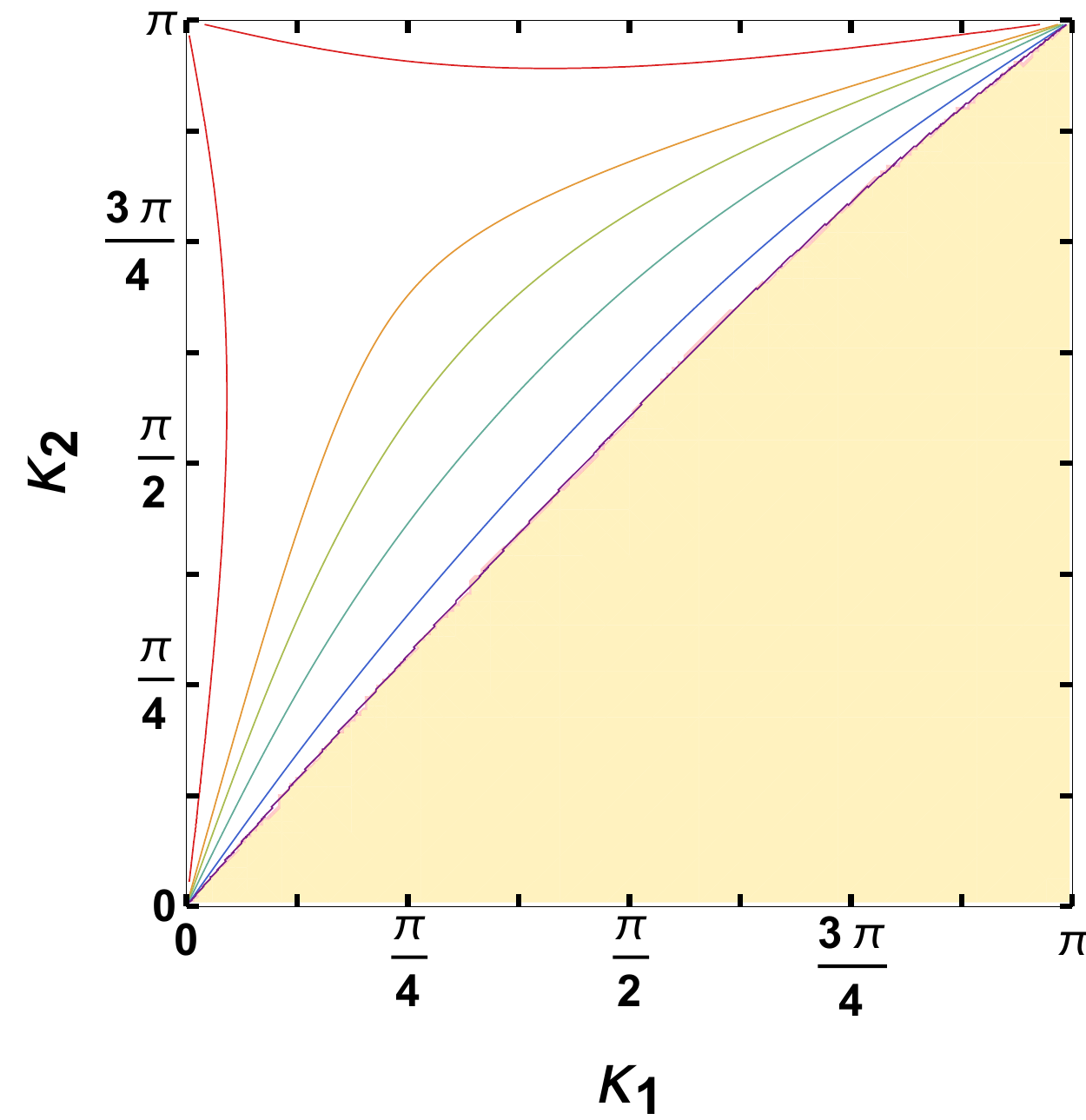} %
\includegraphics[scale=0.5]{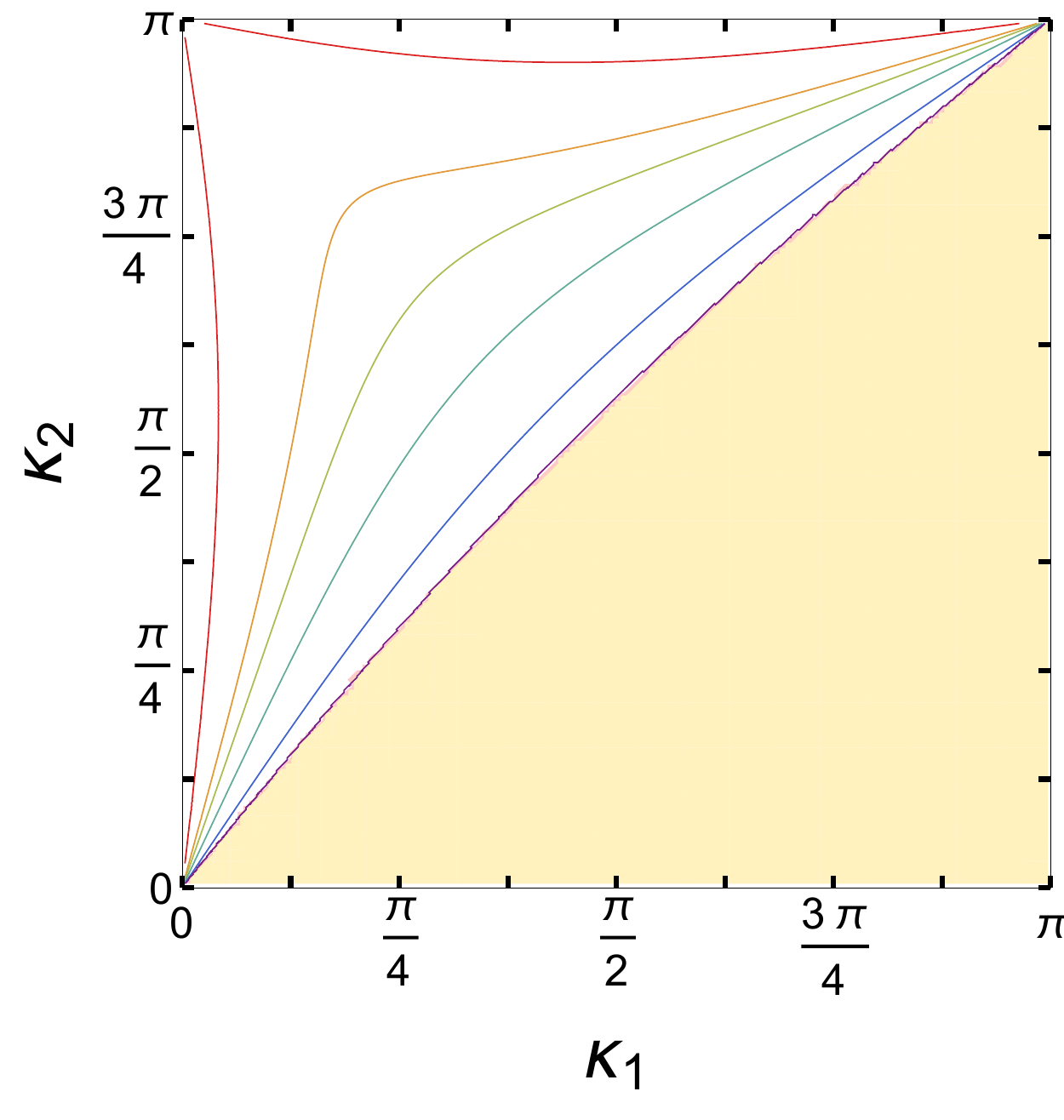} %
\includegraphics[scale=0.5]{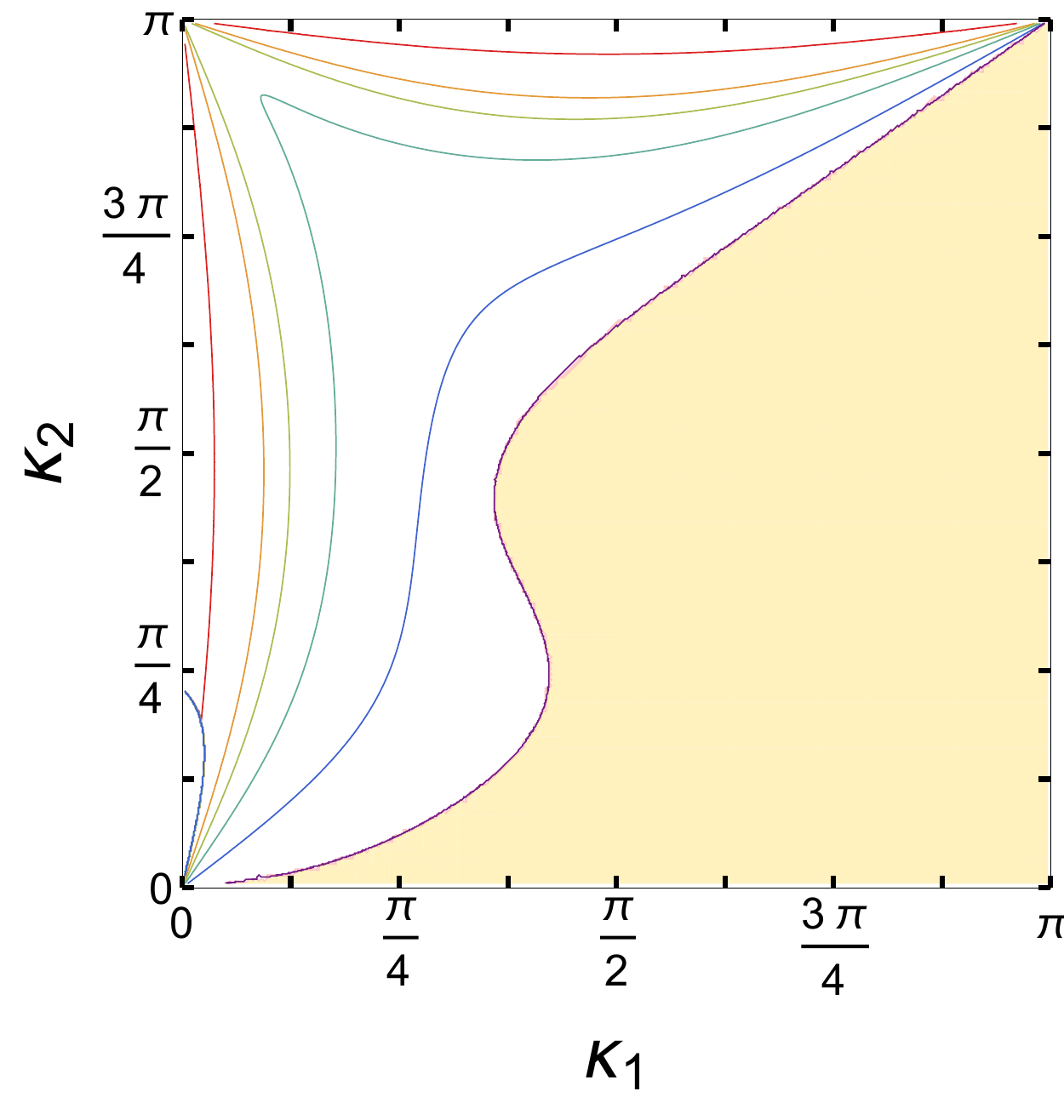} %
\includegraphics[scale=0.5]{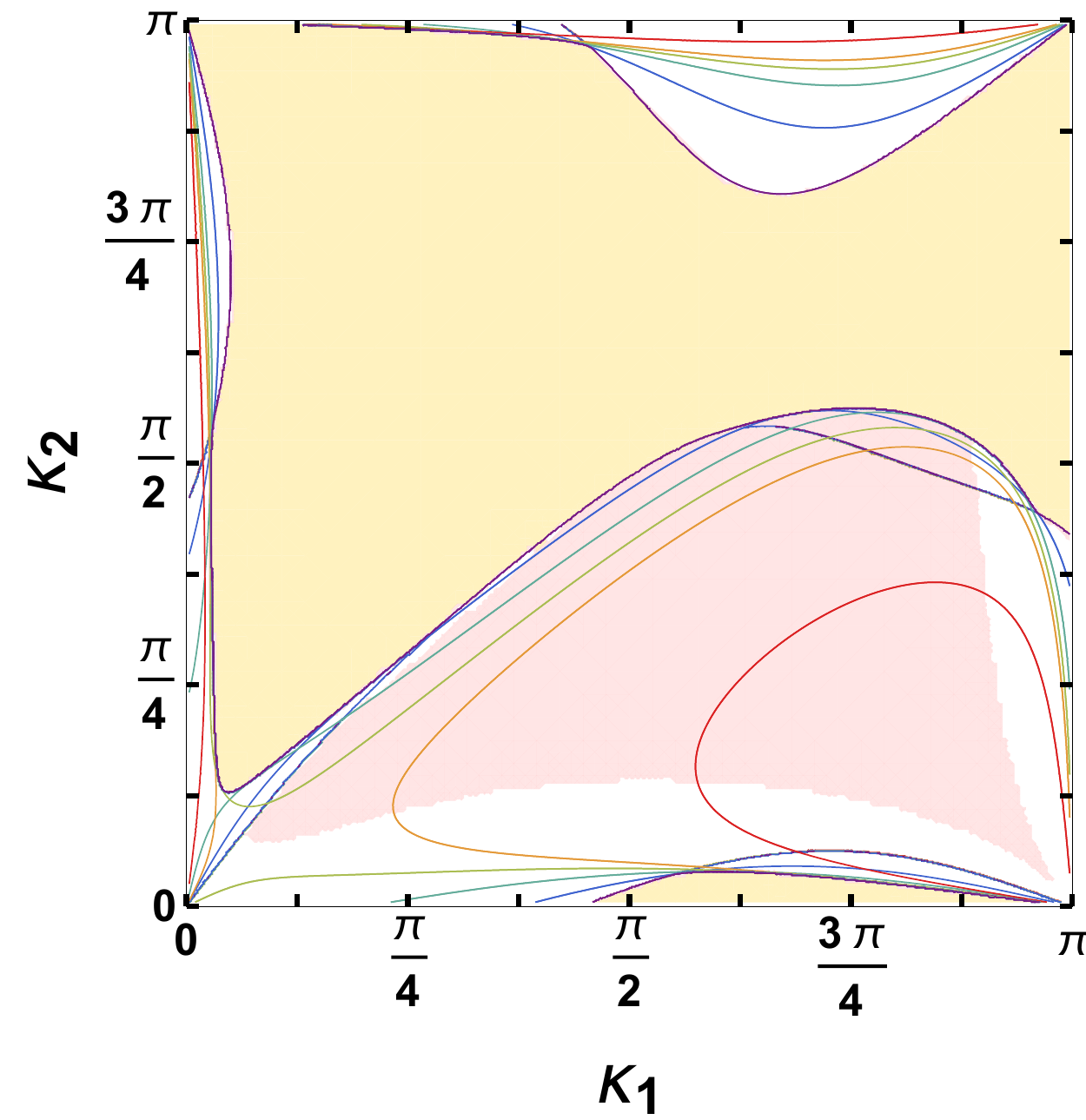}
\end{center}
\caption{Linear stability of the coplanar fixed point configurations with $%
\Delta \protect\zeta _{\left( 0\right) }=0$ as function of $\protect\kappa %
_{\left( 0\right) 1}$ and $\protect\kappa _{\left( 0\right) 2}$, represented
for $\protect\chi _{i}=1$ and $\protect\nu \approx 0.82$. White regions are
marginally stable, and pastel pink are unstable. Yellow regions do not have
fixed points. The purple, blue, green, olive, orange, and red contour lines
refer to the $2$, $4$, $8$, $12$, $16$, and $40$ values of $\mathfrak{\bar{l}%
}_{r}$, respectively. The panels show binaries composed of gravastar -
gravastar with $w_{i}=0$ (top left), black hole - black hole with $w_{i}=1$
(top right), neutron star - neutron star with $m_{NS_{1}}=1.7M_{\odot }$, $%
w_{NS_{1}}=2.55$, $m_{NS_{2}}=1.4M_{\odot }$, $w_{NS_{1}}=4.3$ (bottom
left), and boson star - boson star with $w_{BS_{1}}=17$, $w_{BS_{2}}=22$,
respectively. }
\label{epsDzero}
\end{figure*}

\subsection{Early inspiral limit}

At sufficiently large separations, where $w_{1}x_{1}/\nu \ll 1$ and $\nu
w_{2}x_{2}\ll 1$ hold, $\omega ^{2}$ reduces to%
\begin{eqnarray}
\frac{\mathfrak{\bar{l}}_{r}^{2}\omega ^{2}}{R^{2}} &=&\left[ 3\left( 2+\nu
^{-1}+\nu \right) -\left( 1+\nu \right) w_{1}\right.  \notag \\
&&\left. -\left( 1+\nu ^{-1}\right) w_{2}\right] \chi _{1}\chi _{2}\sin
\kappa _{\left( 0\right) 1}\sin \kappa _{\left( 0\right) 2}  \notag \\
&&+2\epsilon _{\Delta \zeta }\left( 2+\nu ^{-1}+\nu \right) \chi _{1}\chi
_{2}\cos \kappa _{\left( 0\right) 1}\cos \kappa _{\left( 0\right) 2}  \notag
\\
&&+\epsilon _{\Delta \zeta }\left( 1+\nu ^{-1}\right) ^{2}\chi _{1}^{2}\frac{%
\sin ^{2}\kappa _{\left( 0\right) 1}}{\sin ^{2}\kappa _{\left( 0\right) 2}} 
\notag \\
&&+\epsilon _{\Delta \zeta }\left( 1+\nu \right) ^{2}\chi _{2}^{2}\frac{\sin
^{2}\kappa _{\left( 0\right) 2}}{\sin ^{2}\kappa _{\left( 0\right) 1}}~.
\end{eqnarray}%
Note that $\mathfrak{\bar{l}}_{r}^{2}$ has factored out. For $\chi _{2}/\chi
_{1}\ll 1$, there is stability for $\Delta \zeta _{\left( 0\right) }=0$ and
instability for $\Delta \zeta _{\left( 0\right) }=\pi $. For generic spins,
but $w_{1}=w_{2}=w$ and $sgn\left( \kappa _{\left( 0\right) 1}\right)
=sgn\left( \kappa _{\left( 0\right) 2}\right) $, there is stability for $%
\Delta \zeta _{\left( 0\right) }=0$ if $w\leq 3$ (including gravastars,
black holes, and some of the neutron stars) and instability for $\Delta
\zeta _{\left( 0\right) }=\pi $ if $w\geq 3~$(the complementary set of
neutron stars and boson stars).

\subsection{Numerical investigation for high spins}

\begin{figure*}[th]
\begin{center}
\includegraphics[scale=0.5]{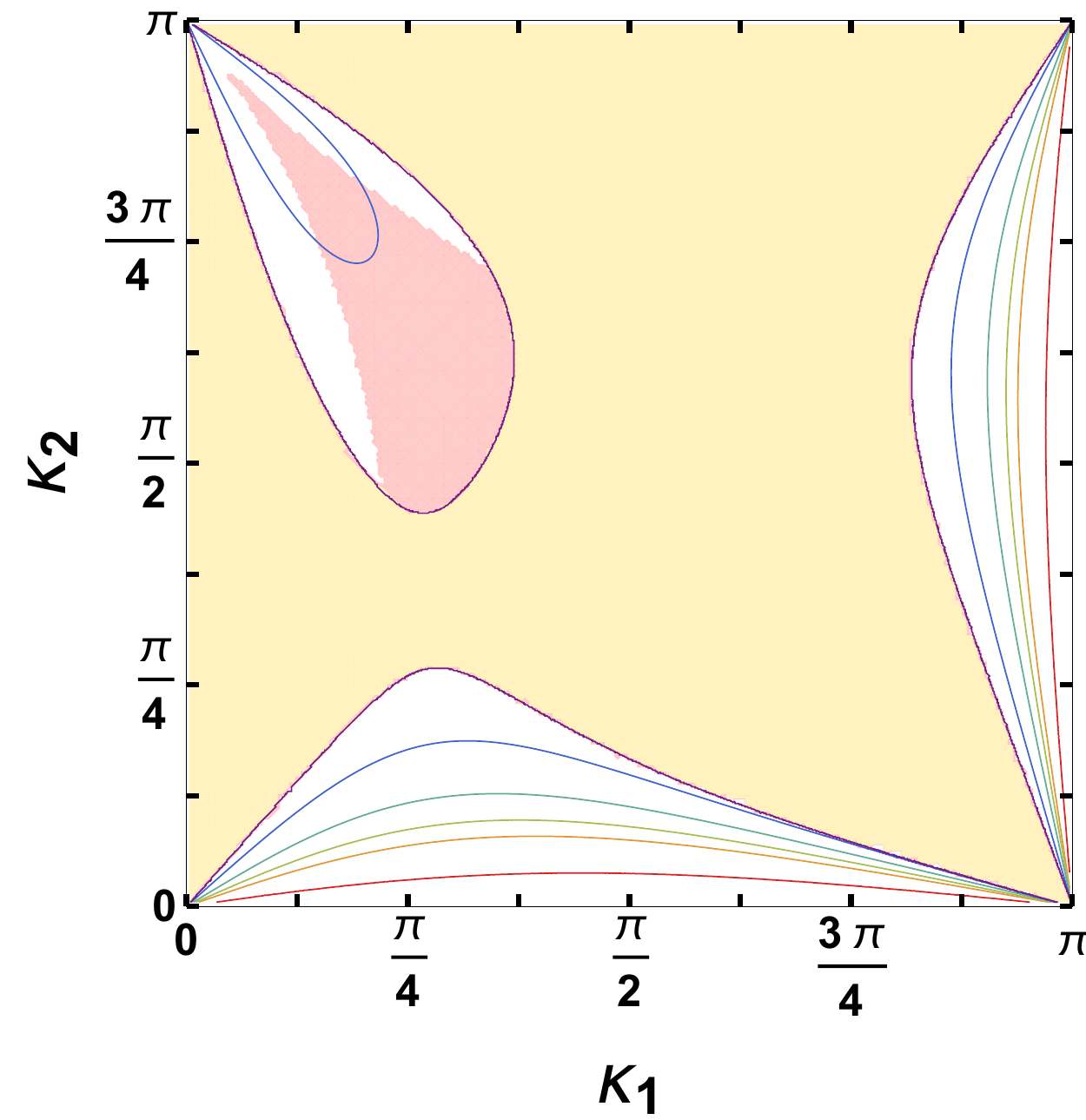} %
\includegraphics[scale=0.5]{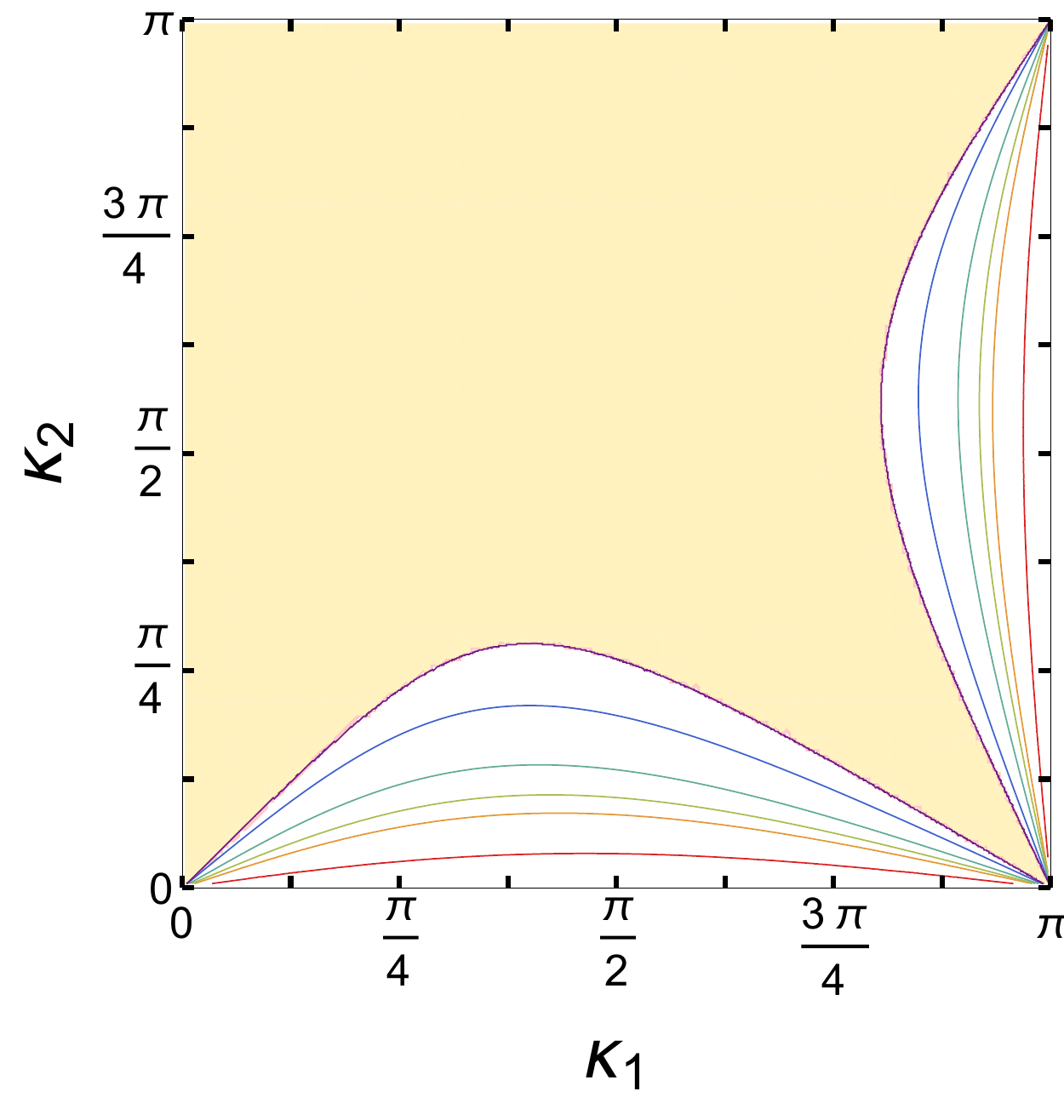} %
\includegraphics[scale=0.5]{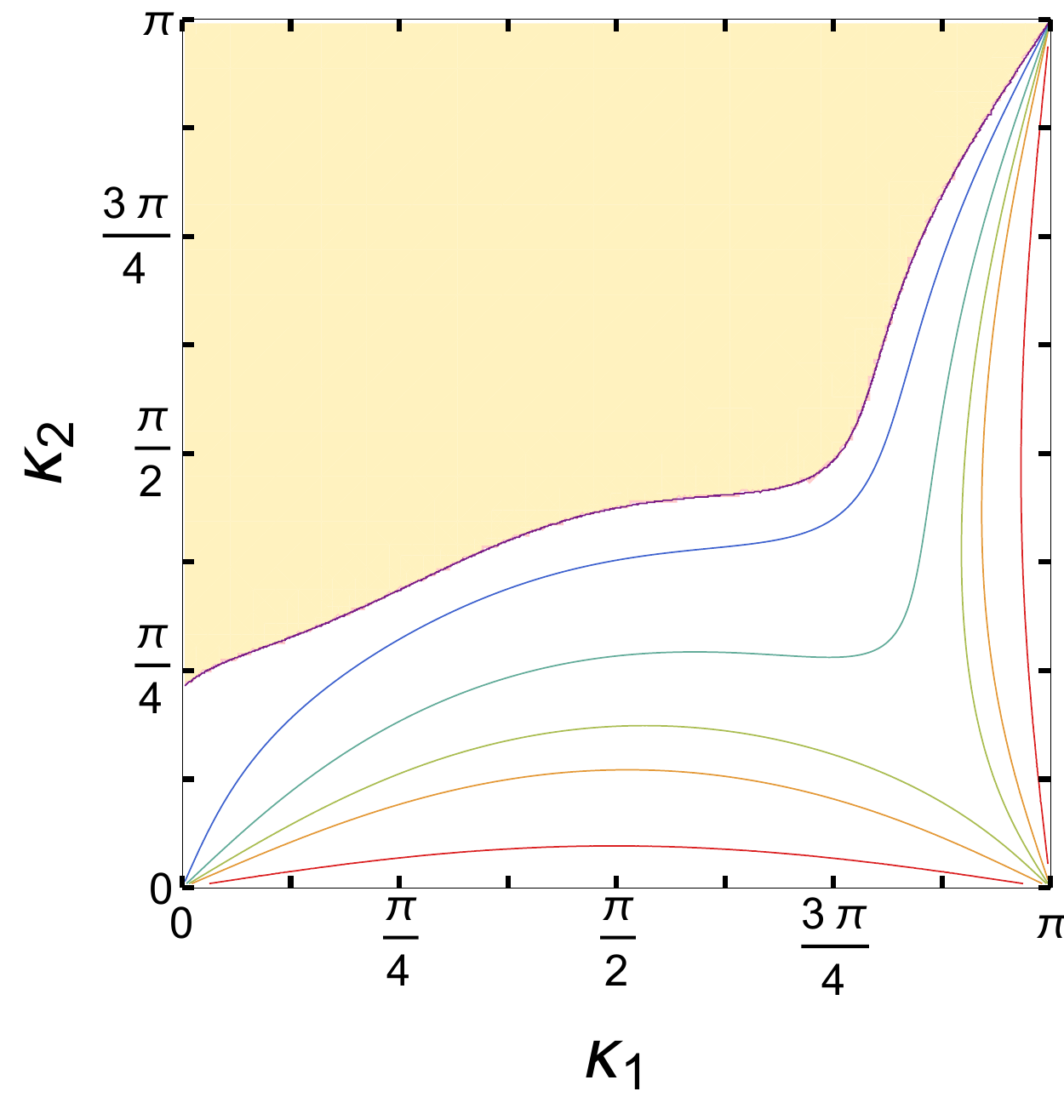} %
\includegraphics[scale=0.5]{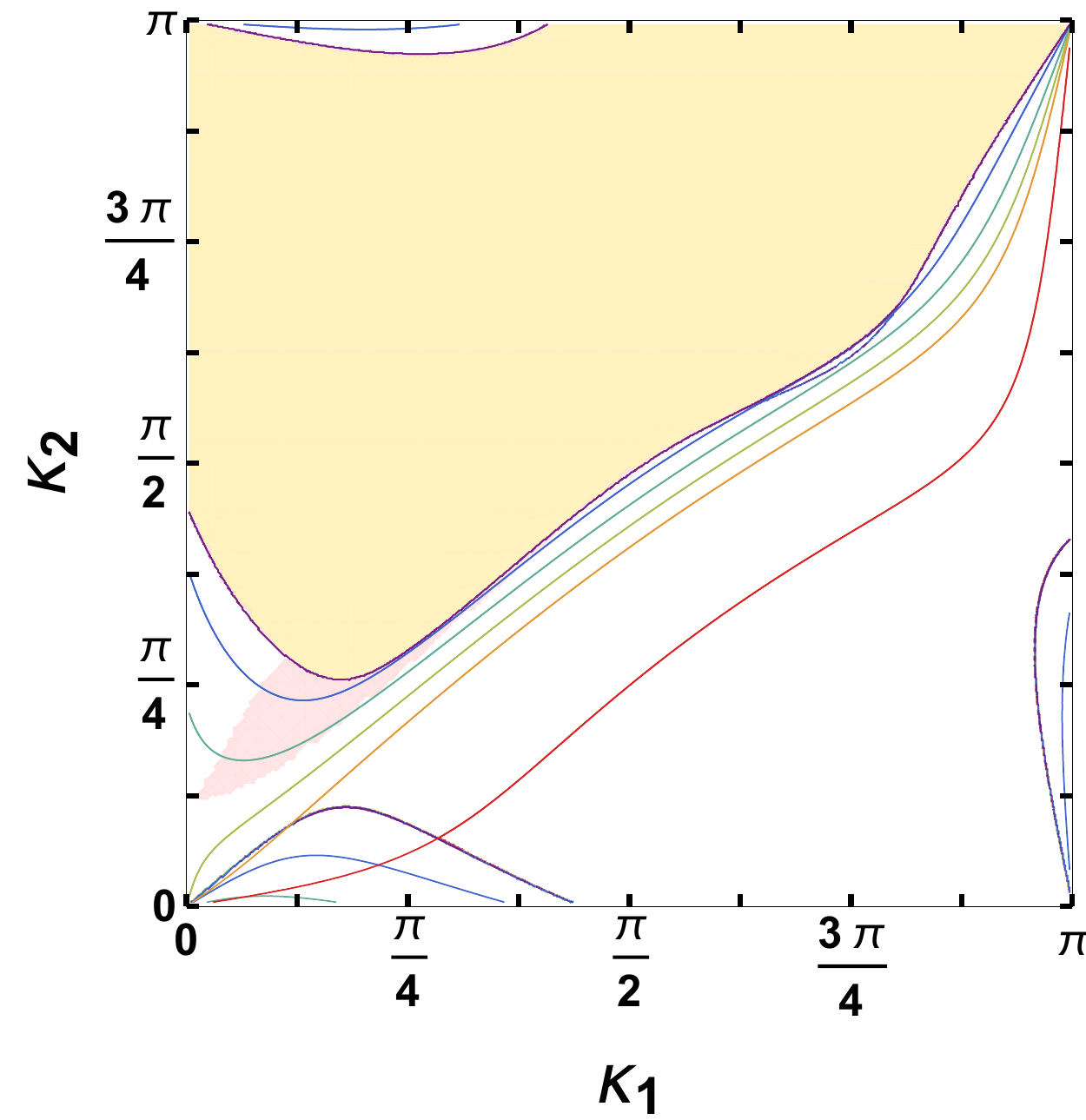}
\end{center}
\caption{Linear stability analysis for the coplanar configuration with $%
\Delta \protect\zeta _{\left( 0\right) }=\protect\pi $. The panels and color
codes are identical to those of Fig. \protect\ref{epsDzero}.}
\label{epsDpi}
\end{figure*}
\begin{figure*}[th]
\begin{center}
\includegraphics[scale=0.45]{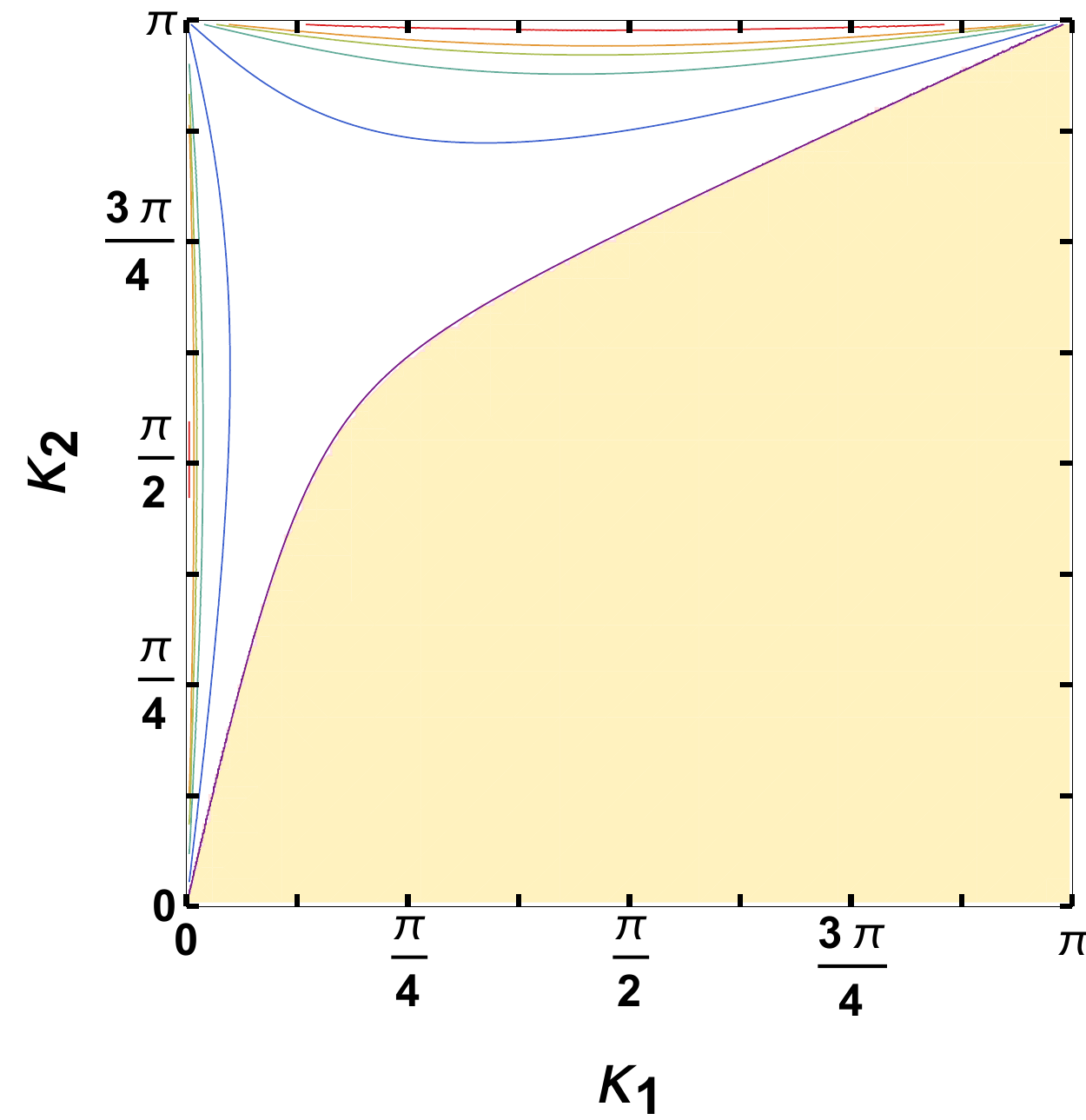} %
\includegraphics[scale=0.45]{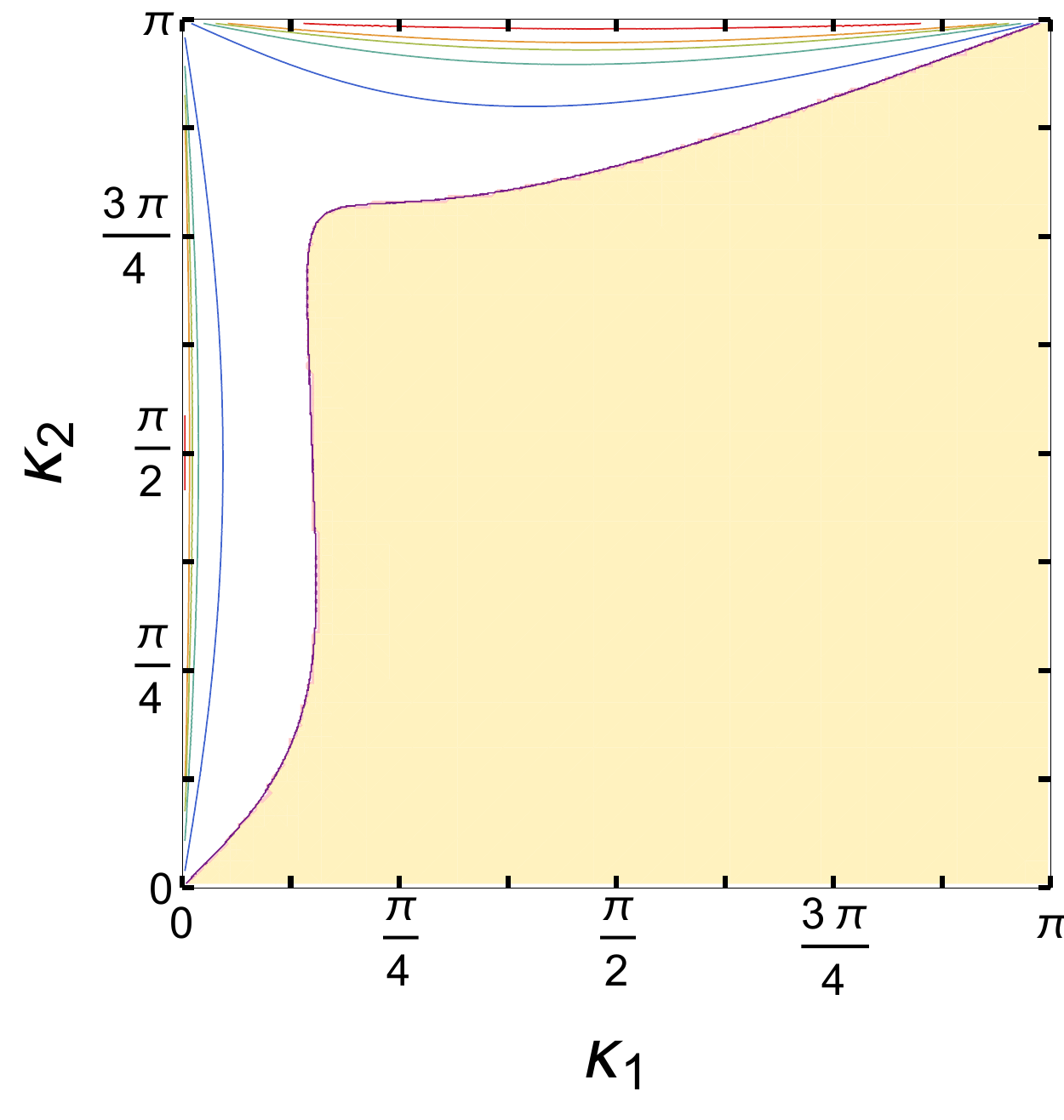} %
\includegraphics[scale=0.45]{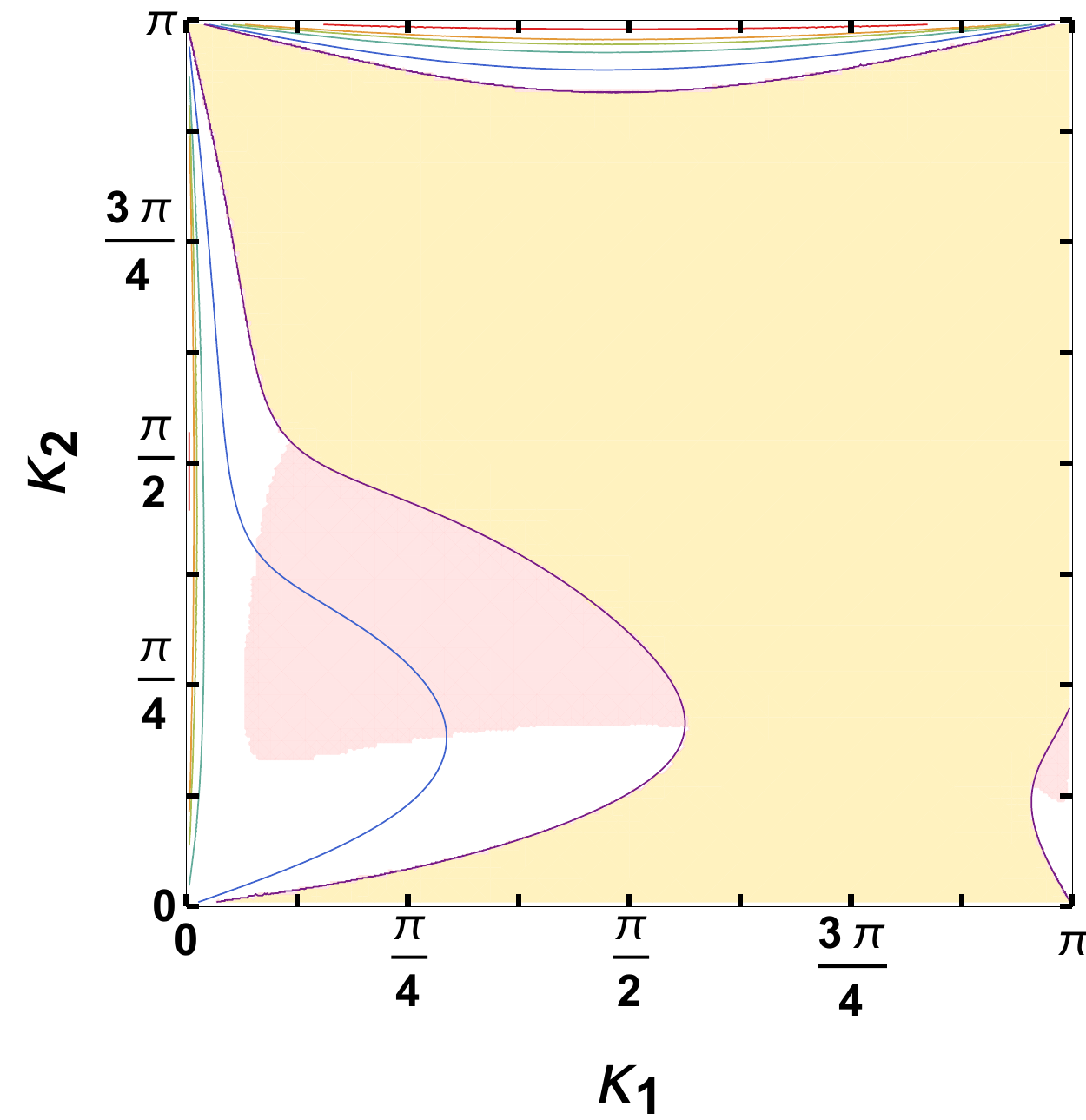}
\end{center}
\caption{Linear stability of the coplanar fixed point configurations with $%
\Delta \protect\zeta _{\left( 0\right) }=0$ as function of $\protect\kappa %
_{\left( 0\right) 1}$ and $\protect\kappa _{\left( 0\right) 2}$, represented
for $\protect\chi _{i}=1$ and $\protect\nu \approx 0.82$. White regions are
marginally stable, and pastel pink are unstable. Yellow regions do not have
fixed points. The purple, blue, green, olive, orange, and red contour lines
refer to the $2$, $4$, $8$, $12$, $16$, and $40$ values of $\mathfrak{\bar{l}%
}_{r}$, respectively. The panels show binaries composed of black hole -
gravastar with $w_{GS}=0$ (left), black hole - neutron star with $%
m_{NS}=1.4M_{\odot }$, $w_{NS}=4.3$ (middle), and black hole - boson star
with $w_{BS}=17$, respectively. }
\label{diffsepsDzero}
\end{figure*}
\begin{figure*}[th]
\begin{center}
\includegraphics[scale=0.45]{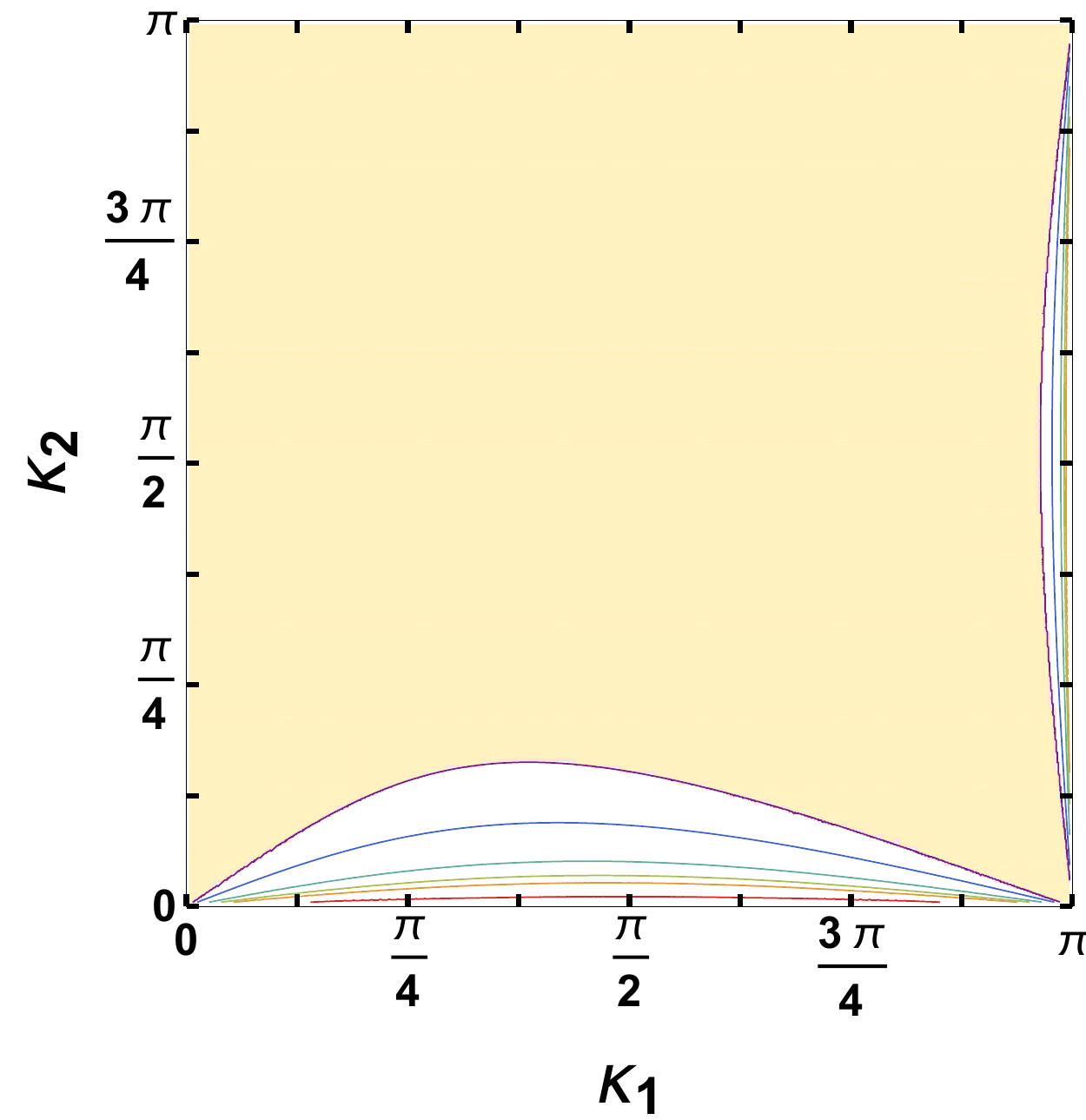} %
\includegraphics[scale=0.45]{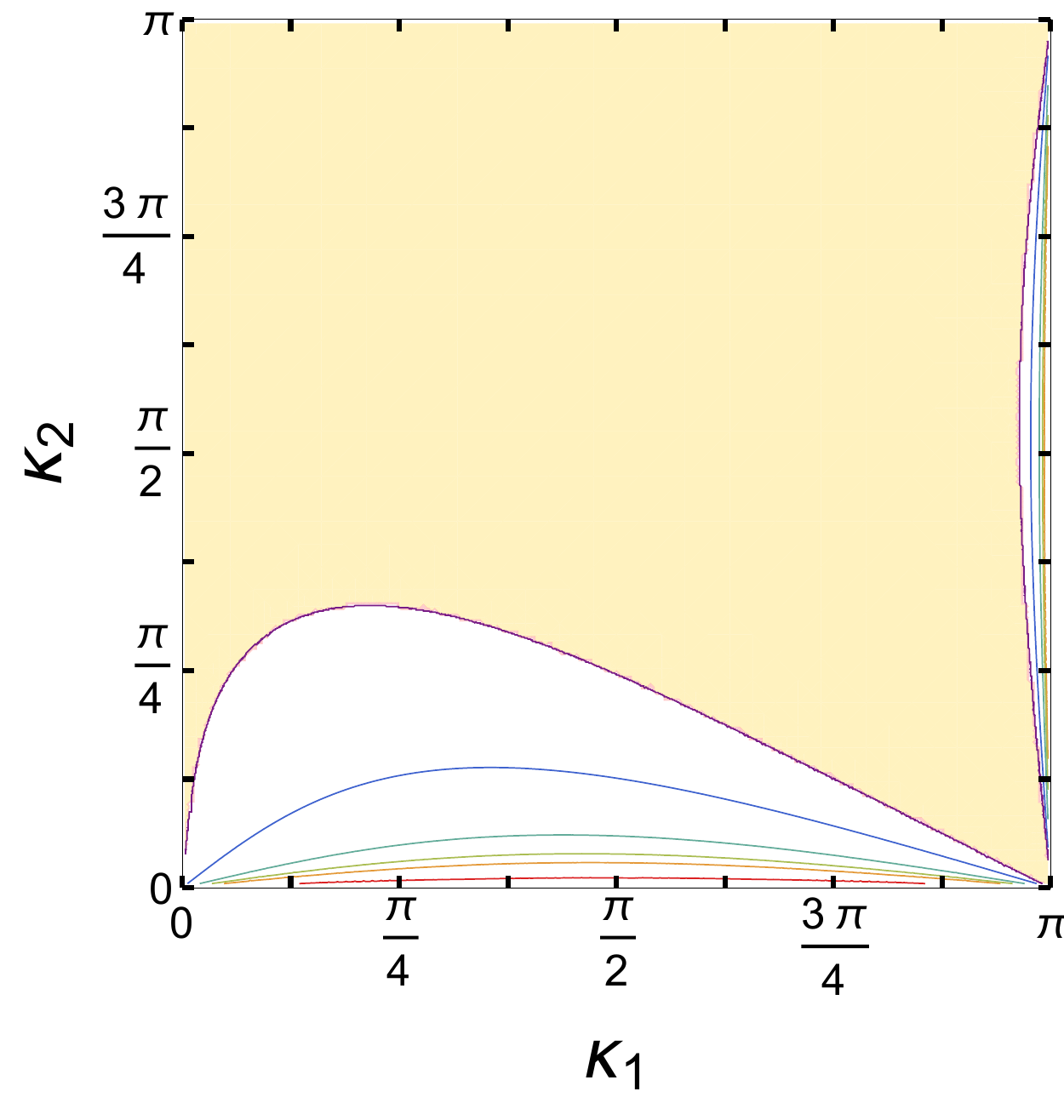} %
\includegraphics[scale=0.45]{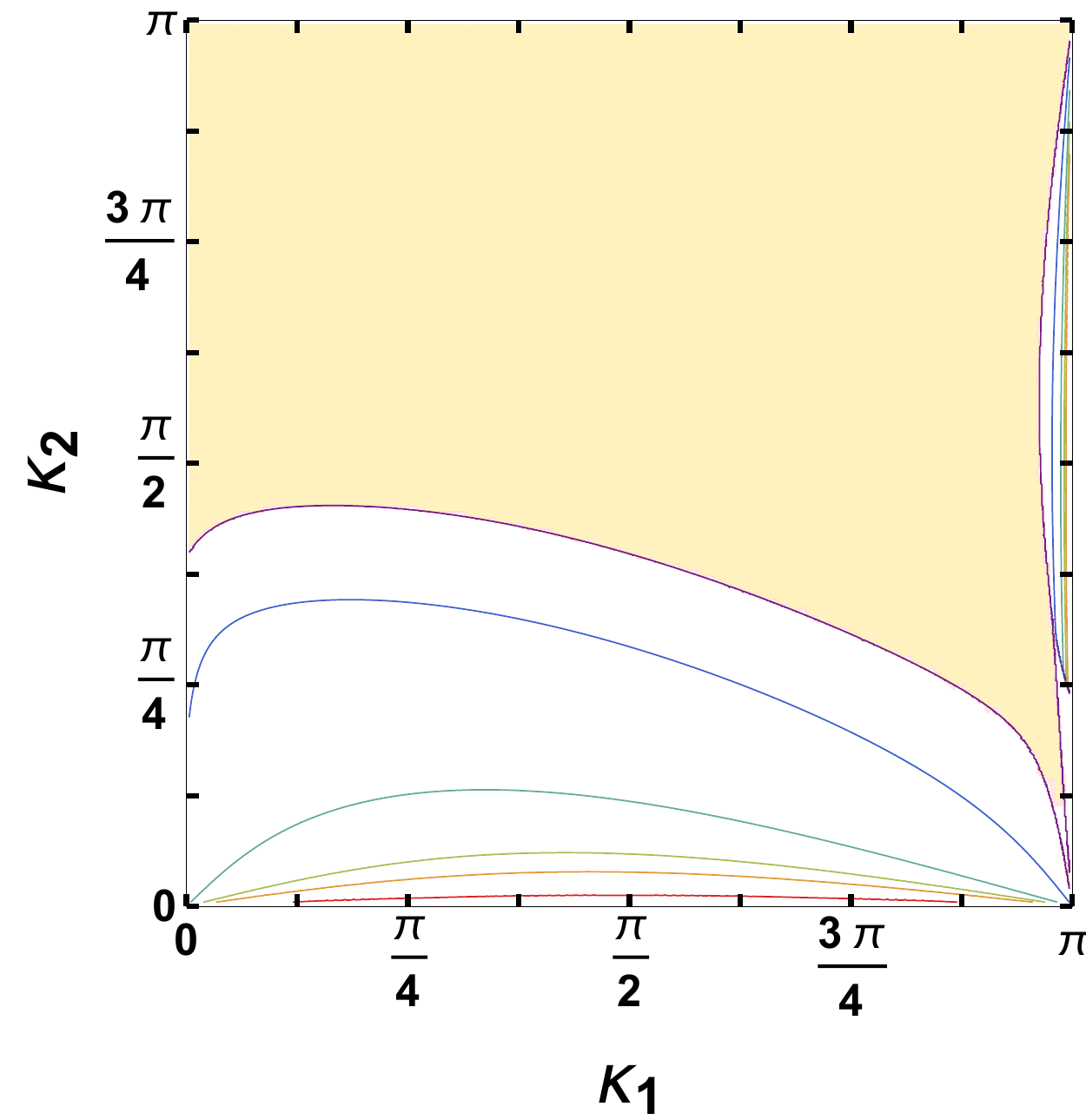}
\end{center}
\caption{Linear stability analysis for the coplanar configuration with $%
\Delta \protect\zeta _{\left( 0\right) }=\protect\pi $. The panels and color
codes are identical to those of Fig. \protect\ref{diffsepsDzero}.}
\label{diffsepsDpi}
\end{figure*}

The stability of coplanar fixed points emerging as solutions of the
constraint (\ref{constr}) depends on the sign of $\omega ^{2}$. Both the
location and number of fixed points and their stability depend severely on
the parameters (\ref{params}). By fixing $\nu $,$~x_{i}$,$~w_{i}$, and $%
\epsilon _{\Delta \zeta }$ the constraint (\ref{constr}) reduces to curves
in the parameter plane of spin polar angles $\kappa _{\left( 0\right) i}$.

Instead of $\,x_{i}$ we may also fix the spin magnitudes $\chi _{i}$, but
then $\mathfrak{\bar{l}}_{r}$ emerges as a third parameter. We represent
then the constraint as contour curves of constant $\mathfrak{\bar{l}}_{r}$
in the parameter plane $\left( \kappa _{\left( 0\right) 1},\kappa _{\left(
0\right) 2}\right) $.

By Eq. (8) of Ref. \cite{chameleon} to leading order, $\mathfrak{\bar{l}}%
_{r}^{2}=\left( 1-\bar{e}_{r}^{2}\right) /\bar{\varepsilon}$. For $\bar{%
\varepsilon}<0.1$, on circular orbits $\mathfrak{\bar{l}}_{r}>3$. For a
nonvanishing eccentricity, $\bar{e}_{r}^{2}<0.75$, the limit $\mathfrak{\bar{%
l}}_{r}>2$ still holds. We do not discuss higher eccentricities, as
gravitational radiation is supposed to consume it efficiently toward the end
of the inspiral. This implies that in Figs. \ref{epsDzero}-\ref{diffsepsDpi}
we will visualize contour curves for $\mathfrak{\bar{l}}_{r}>2$ only.

For Fig. \ref{epsDzero}, $\Delta \zeta _{\left( 0\right) }=0$, $\chi _{i}=1$%
, and $\nu \approx 0.82$ were chosen. The first panel, representing
gravastar binaries with $w_{i}=0$ reproduces the contour lines from Fig. 2
of Ref. \cite{Schnittman}, where the mass quadrupole was disregarded. The
purple, blue, green, olive, orange and red contour lines emerge for the
values $2$ ($0.5$), $4$ ($1$), $8$ ($2$), $12$ ($3$), $16$ ($4$), and $40$ of%
$\ \mathfrak{\bar{l}}_{r}$. The numbers in brackets (where applicable) are
the corresponding $L_{N}$, given for easier comparison with Fig. 2 of Ref. 
\cite{Schnittman}. In the white region the fixed points are marginally
stable (with any deviation leading to libration) while in the yellow area,
there are no fixed points (the deviations increase monotonically). The
second panel represents black hole binaries, i.e., $w_{i}=1$, with the same
color codes and contour lines. They are slightly deformed as compared to the
gravastar binary. The third panel represents a neutron star - neutron star
binary with masses $m_{NS_{1}}=1.7M_{\odot }$ and $m_{NS_{2}}=1.4M_{\odot }$%
. According to Table VII. of Ref. \cite{NSw} the respective quadrupole
parameters of the neutron stars with a FPS (Friedman-Pandharipande + Skyrme)
equation of state are $w_{NS_{1}}=2.55$ and $w_{NS_{2}}=4.3$. The fourth
panel represents a boson star - boson star binary with quadrupole parameters 
$w_{BS_{1}}=17$ and $w_{BS_{2}}=22$. These quadrupole parameters roughly
correspond to the values on the bottom two curves of Fig. 4 in Ref. \cite%
{BosonStarw} at $\chi _{i}=1$. These two curves represent a system with mass
ratio $\nu =0.82$, as on our figure. The contour lines of $\mathfrak{\bar{l}}%
_{r}$ cross each other in certain regions for the boson star - boson star
system. However, this is no problem, as the crossing lines correspond to
different values of $\mathfrak{\bar{l}}_{r}$ and thus to distinct phases of
the inspiral. It is clear from a comparison of the four panels that the
contour lines of $\mathfrak{\bar{l}}_{r}$ depend significantly on the
quadrupole parameter. They run on a white background, representing the
parameter regions with linear stability. The red patch on the fourth panel
represents the unstable parameter region. Note the contour lines which
extend on both stable and unstable regions. On the yellow background, there
are no fixed points.

In the four panels of Fig. \ref{epsDpi}, the contour lines for all four
above-mentioned binary systems are shown for $\Delta \zeta _{\left( 0\right)
}=\pi $, all other parameters identical. In this case, the value of the
quadrupole parameter affects even more the stability analysis. Unstable
regions appear both for the gravastar - gravastar and boson star - boson
star binaries. The stability region is much larger for neutron stars than
black holes. Overlapping contour curves appear only for the boson star
binaries.

The regions which contain unstable patches embedded in a linearly stable
region deserve additional investigation. We have checked numerically that
whenever linear stability holds in these regions generic nonlinear
perturbations confirm the stability up to perturbations of $0.5\%$ but can
destroy stability if the perturbation is roughly larger than $1\%$. This
happens as second- and higher order perturbative effects become important.
This feature is in agreement with the analysis of Ref. \cite{Schnittman}
intended to be performed for black holes, but as the mass quadrupole was
disregarded, they rather hold for gravastars with $w_{i}=0$).

In Fig. \ref{diffsepsDzero} $\Delta \zeta _{\left( 0\right) }=0$, $\chi
_{i}=1$ and $\nu \approx 0.28$ were chosen. The first, second and third
panels represent black hole - gravastar, black hole - neutron star, and
black hole - boson star binaries, respectively. The quadrupole parameters
for the gravastar, neutrons star, and boson star are $w_{GS}=0$, $w_{NS}=4.3$
(corresponding to the mass $m_{NS}=1.4M_{\odot }$) and $w_{BS}=17$. These
systems are also visualized in Fig. \ref{diffsepsDpi} for $\Delta \zeta
_{\left( 0\right) }=\pi $. As a generic rule the stability regions are
smaller than for binaries of identical nature.

\begin{table*}[tbp]
\begin{center}
\begin{tabular}{|c||c|c|c|}
\hline\hline
$\uparrow L_{N}$ & $\uparrow S_{1}\quad \uparrow S_{2}$ & $\uparrow
S_{i}\quad \downarrow S_{j\neq i}$ & $\downarrow S_{1}\quad \downarrow S_{2}$
\\ \hline\hline
\begin{tabular}{c}
\\ 
$\nu \ll 1$%
\end{tabular}
& 
\begin{tabular}{c}
\\ 
$+$%
\end{tabular}
& 
\begin{tabular}{c}
\\ 
$+$%
\end{tabular}
& 
\begin{tabular}{c}
\\ 
$+$%
\end{tabular}
\\ \hline
\begin{tabular}{c}
$\nu \lesssim 1,$ \\ 
$\chi _{1}=\chi _{2}=\chi ,$ \\ 
$w_{1}=w_{2}=w$%
\end{tabular}
& 
\begin{tabular}{c}
$+$ \\ 
(illustrated for \\ 
$x\equiv \frac{\chi }{\bar{l}_{r}}=0.3$ \\ 
in Fig. \ref{corotMS})%
\end{tabular}
& 
\begin{tabular}{c}
\\ 
$+$ (BGS$\downarrow \uparrow $, BBH$\downarrow \uparrow $, \\ 
BNS$\uparrow \downarrow $hi$w$, BBS$\uparrow \downarrow $) \\ \hline
\\ 
Evolves during the inspiral as \\ 
$+-$ (BGS$\uparrow \downarrow $hi$w$, BBH$\uparrow \downarrow $, BNS$%
\uparrow \downarrow $lo$w$, BNS$\downarrow \uparrow $lo$w$) \\ \hline
\\ 
Evolves during the inspiral as \\ 
$+-+$ (BGS$\uparrow \downarrow $lo$w$, BNS$\downarrow \uparrow $hi$w$, BBS$%
\downarrow \uparrow $) \\ 
(shown for $\nu =0.9$ in Fig. \ref{Contour1}, \\ 
for arbitrary $\nu $ in Fig. \ref{noneqmass})%
\end{tabular}
& $+$ \\ \hline
\begin{tabular}{c}
$\nu =1,$ \\ 
$\chi _{1}=\chi _{2}=\chi ,$ \\ 
$w_{1}=w_{2}=w$%
\end{tabular}
& 
\begin{tabular}{c}
$+$ for BGS, BBH \\ \hline
\\ 
$+$ for BNS and BBS if$\ w\neq w_{cr1}$, \\ 
where $w_{cr_{1}}=\frac{2}{x}-1$ $>\frac{17}{3}$ \\ 
decreases during the inspiral%
\end{tabular}
& 
\begin{tabular}{c}
\\ 
$+$ for BGS, BBS \\ \hline
\\ 
$+$ for BNS with $w>w_{+}$, \\ 
where $w_{+}\equiv 1+\frac{4}{x^{2}+1}\in \left( 4.7,5\right) $ \\ 
decreases during the inspiral \\ \hline
\\ 
$-$ for BBH, BNS with $w<w_{+}$%
\end{tabular}
& $+$ \\ \hline
\begin{tabular}{c}
Sufficient \\ 
conditions \\ 
in all \\ 
other cases%
\end{tabular}
& 
\begin{tabular}{c}
\\ 
$+$ for BGS, BBH, GS-BH \\ \hline
\\ 
$+$ for BNS with $w_{1}<\frac{10}{3}$, $w_{2}<\frac{17}{3}$ \\ \hline
\\ 
$+$ for BNS or BBS \\ 
with $w_{i}\leq W_{1}^{-}$ and $w_{3-i}<W_{3-i}^{-}$ \\ 
or with $w_{i}\geq W_{1}^{-}$ and $w_{2}>W_{2}^{-}$, \\ 
where $W_{i}^{-}$ decrease during inspiral \\ 
according to Eq. (\ref{Wi}), \\ 
bound as $W_{1}^{-}>\frac{10}{3}$, $W_{2}^{-}>\frac{17}{3}$%
\end{tabular}
& 
\begin{tabular}{c}
$+$ for binaries with a NS or BS \\ 
as the $i^{th}$ component, \\ 
obeying $w_{i}>W_{i}^{+}$, \\ 
where $W_{i}^{+}$ decrease during inspiral \\ 
according to Eq. (\ref{Wi}), \\ 
bound as $W_{1}^{+}>\frac{10}{3}$, $W_{2}^{+}>\frac{20}{3}$%
\end{tabular}
& $+$ \\ \hline
\end{tabular}%
\end{center}
\caption{Summary of results on marginal stability ($+$) or instability ($-$)
of the collinear spin configurations in compact binary systems (B) composed
of gravastars (GS), black holes (BH), neutron stars (NS), and boson stars
(BS) for various mass ratios $\protect\nu =m_{2}/m_{1}\leq 1$ and spin
orientations. The first spin in the name of a binary (for example BBH$%
\downarrow \uparrow $) refers to the higher mass component. The stability
may be different for lower values (lo$w$) or higher values (hi$w$) of the
allowed quadrupolar parameter range [$w_{i}\in $ $\left( -0.8,1\right) $ for
GS, $w_{i}=1$ for BH, $w_{i}\in $ $\left( 2,14\right) $ for NS, and $%
w_{i}\in \left( 10,150\right) $ for BS]. For example BNS$\uparrow \downarrow 
$lo$w$ denotes a binary system of neutron stars with the spin of the more
(less) massive neutron star parallel (antiparallel) to the Newtonian angular
momentum $\mathbf{L}_{\mathbf{N}}$ and with quadrupolar parameter in the
lower range $w\gtrsim 2$. }
\label{tablesum1}
\end{table*}

\begin{table*}[tbp]
\begin{center}
\begin{tabular}{|c||c|c|}
\hline\hline
$\uparrow L_{N}$ & $\uparrow S_{1}$ & $\downarrow S_{1}$ \\ \hline\hline
\begin{tabular}{c}
$\nu =1,$ \\ 
$\chi _{2}=0$%
\end{tabular}
& 
\begin{tabular}{c}
\\ 
$+$ for all type of binaries, \\ 
unless $m_{1}\,$\ is a NS with \\ 
$w_{1}=w_{cr2}^{+}=\frac{3}{1+x_{1}}\in \left( 2.3,3\right) $%
\end{tabular}
& 
\begin{tabular}{c}
\\ 
$+$ for all type of binaries, \\ 
unless $m_{1}\,$\ is a NS with \\ 
$w_{1}=w_{cr2}^{-}=\frac{3}{1-x_{1}}\in \left( 3,4.3\right) $%
\end{tabular}
\\ \hline
\begin{tabular}{c}
$\nu \lesssim 1,$ \\ 
$\chi _{2}=0$%
\end{tabular}
& 
\begin{tabular}{c}
\\ 
$+$ for all type of binaries, \\ 
unless $m_{1}\,$\ is a NS or BS with \\ 
$w_{1}=\frac{\left( 1+2\nu ^{-1}\right) x_{1}-\left( \nu ^{-1}-\nu \right) }{%
\left( \nu ^{-1}x_{1}+1\right) x_{1}}\leq w_{cr2}^{+}$%
\end{tabular}
& 
\begin{tabular}{c}
\\ 
$+$ for all type of binaries, \\ 
unless $m_{1}\,$\ is a NS or BS with \\ 
$w_{1}=-\frac{\left( 1+2\nu ^{-1}\right) x_{1}+\left( \nu ^{-1}-\nu \right) 
}{\left( \nu ^{-1}x_{1}-1\right) x_{1}}\geq w_{cr2}^{-}$%
\end{tabular}
\\ \hline
\begin{tabular}{c}
$\nu \ll 1,$ \\ 
$\chi _{2}=0$%
\end{tabular}
& 
\begin{tabular}{c}
\\ 
$+$ for all type of binaries, \\ 
(no instability in the \\ 
physical range of $w_{1}$)%
\end{tabular}
& 
\begin{tabular}{c}
\\ 
$+$ for all type of binaries, \\ 
(no instability in the \\ 
physical range of $w_{1}$)%
\end{tabular}
\\ \hline
\end{tabular}%
\end{center}
\caption{Marginal stability ($+$) holds generically, when one of the spins
is collinear with the Newtonian angular momentum $\mathbf{L}_{\mathbf{N}}$,
while the second spin is negligible, with the single exception of a
particular value of the quadrupolar parameter of a spinning neutron star
(NS) or boson star (BS) in a binary with $\protect\nu \lesssim 1$.}
\label{tablesum2}
\end{table*}

\begin{table*}[tbp]
\begin{center}
\begin{tabular}{|c||c||c|c|c|c|c|c|c|}
\hline\hline
$\left( \kappa _{1},\kappa _{2}\right) $ & $\Delta \zeta $ & GSB & BHB & NSB
& BSB & BH-GS & BH-NS & BH-BS \\ \hline\hline
$\left( \frac{3\pi }{8},\frac{\pi }{4}\right) $ & $0$ & $\diagup $ & $%
\diagup $ & $+$ & $-$ & $\diagup $ & $\diagup $ & $-$ \\ \cline{2-9}
& $\pi $ & $\diagup $ & $+$ & $+$ & $+$ & $\diagup $ & $+$ & $+$ \\ 
\hline\hline
$\left( \frac{3\pi }{8},\frac{3\pi }{4}\right) $ & $0$ & $+$ & $+$ & $+$ & $%
\diagup $ & $+$ & $\diagup $ & $\diagup $ \\ \cline{2-9}
& $\pi $ & $-$ & $\diagup $ & $\diagup $ & $\diagup $ & $\diagup $ & $%
\diagup $ & $\diagup $ \\ \hline
\end{tabular}%
\end{center}
\caption{Regions with marginal stability ($+$), instability ($-$), and with
no fixed points ($\diagup $) of a given spin configuration $\left( \protect%
\kappa _{1},\protect\kappa _{2}\right) $ in the coplanar cases $\Delta 
\protect\zeta =0,\protect\pi $ for various binaries. For each binary, the
pair $\left( \protect\kappa _{1},\protect\kappa _{2}\right) $ refers to a
particular phase of the inspiral represented by the corresponding level
curve in Figs. \protect\ref{epsDzero}--\protect\ref{diffsepsDpi}. }
\label{wdef}
\end{table*}

\section{Conclusions\label{concludingr}}

In this paper, we investigated the closed system of first-order differential
equations for the spin orientations derived in Paper I from the dynamical
system analysis point of view, by performing a linear stability analysis of
the evolutions of spin polar angles and the difference of their azimuthal
angles.

By investigating a sufficient condition of stability:

\begin{enumerate}
\item[i)] systems with both spins antialigned with the orbital angular
momentum were found marginally stable.

\item[ii)] systems of black hole - black hole, gravastar - gravastar, and
black hole - gravastar binaries with both spins aligned with the orbital
angular momentum were also found marginally stable. Binaries with neutron
star and boson star components and aligned spins could be marginally stable
in various stages of the inspiral depending on how their quadrupolar
parameter compares to the functions $W_{i}^{-}$ which decrease monotonically
during the inspiral.

\item[iii)] in the configurations with one spin aligned and another
antialigned, marginal stability could hold either during the whole inspiral
for sufficiently large values of $w_{i}$, or only at the latter stages of
the inspiral.
\end{enumerate}

In the analysis of the case of equal masses, spins, and quadrupole
parameters, we investigated a necessary condition for marginal stability.
This has resulted in:

\begin{enumerate}
\item[a)] the confirmation of our earlier finding that the configurations
with the spins antialigned to the orbital angular momentum are marginally
stable.

\item[b)] the aligned configurations also turned out marginally stable, with
the exception of a critical quadrupolar parameter value allowed for neutron
stars and boson stars only.

\item[c)] for black holes and gravastars, the configurations with one spin
aligned and another antialigned to the orbital angular momentum being found
unstable. For neutron stars, these configurations were found either
marginally stable or unstable (depending on the equation of state), while
for boson stars they turned out marginally stable.
\end{enumerate}

These results generalize the stability analysis result performed earlier for
black hole binaries (which had the mass quadrupolar contributions neglected,
hence rather holding for gravastars with $w_{i}=0$\thinspace ).

For nonequal masses, we found that marginal stability occurs for all values
of the quadrupolar parameter (for gravastars, black holes, neutron stars, or
boson stars in binary configurations) when both spins are either antialigned
or aligned to the orbital angular momentum. Other cases were represented in
Fig. \ref{Contour1}.

For black hole binaries, we recovered the transition from stability to
instability during the inspiral occurring when the spin of the larger black
hole is aligned to the orbital angular momentum and that of the smaller mass
black hole is antialigned, discussed in Ref. \cite{GKSKBDT}. In the opposite
alignment case, the evolution was marginally stable.

We identified similar evolutions leading to the abrupt disappearance of
stability for neutron star binaries in the quadrupole parameter range $w\in
\left( 2,4\right) $. Contrary to black hole binaries, this can happen if any
of the neutron star spins is aligned to the orbital angular momentum, while
the other is antialigned. Neutron star binaries with $w>4$, with the larger
mass neutron star spin aligned to the orbital angular momentum, are always
marginally stable.

We also identified configurations allowing for a sequence of evolutions,
which are stable, then unstable, then stable again during the inspiral. In
the following cases, a transitional instability occurs:

\begin{enumerate}
\item[A)] gravastar binaries with $w$ values in the lower part of their
allowed range, in the configuration of the larger mass gravastar spin
aligned to the orbital angular momentum;

\item[B)] neutron star binaries with $w$ values in the higher part of their
allowed range, with the smaller mass neutron star spin aligned to the
orbital angular momentum;

\item[C)] boson star binaries, also in the configuration of the smaller mass
boson star spin aligned to the orbital angular momentum (in this case the
instability occurs only for a brief part of the evolution).
\end{enumerate}

We summarize these results in Table \ref{tablesum1}.

Marginal stability holds generically, when one of the spins is collinear
with the Newtonian angular momentum $\mathbf{L}_{\mathbf{N}}$, while the
second spin is negligible, with the single exception of a particular value
of the quadrupolar parameter of a spinning neutron star or boson star in a
binary with comparable masses. The details are summarized in Table \ref%
{tablesum2}.

Coplanar configurations (of the spin vectors and orbital angular momentum)
also allow for fixed points. We discussed the linear stability of them in
the parameter plane of the spin polar angles, fixing all other parameters
with the exception of the monotonously decreasing $\mathfrak{\bar{l}}_{r}$
during the inspiral, the different values of which were represented as level
curves. We reproduced earlier results holding for gravastar binaries. We
also investigated the linear stability of black hole, neutron star and boson
star binaries, also of mixed black hole - gravastar, black hole - neutron
star, and black hole - boson star binaries. In the particular numerical
examples discussed, we found that instabilities occur only for the gravastar
- gravastar, boson star - boson star, and black hole\ - boson star binaries.

Our analysis highlights that the marginal stability of the spin
configurations strongly depends on the value of the quadrupolar parameter,
as shown in Figs. \ref{epsDzero}--\ref{diffsepsDpi}. We illustrate this
statement in Table \ref{wdef} for four particular coplanar configurations.
Any configuration in the $\left( \kappa _{1},\kappa _{2}\right) $ parameter
plane could be marginally stable only for the corresponding value of $%
\mathfrak{\bar{l}}_{r}$, represented by level curves. Fixing $\mathfrak{\bar{%
l}}_{r}$ (for example the blue curve), in other words picking a particular
phase of the inspiral and the value of $\kappa _{1}$ in Figs. \ref{epsDzero}%
--\ref{diffsepsDpi} allows us to read the value(s) of $\kappa _{2}$ of the
marginally stable configuration at the respective phase of the inspiral.

The stability region is much larger for neutron star binaries than for black
hole binaries in the coplanar configuration. Finally the mixed systems
exhibit a restricted stability parameter region.

\section{Acknowledgements}

This work was supported by the Hungarian National Research Development and
Innovation Office (NKFIH) in the form of Grant No. 123996 and has been
carried out in the framework of COST actions CA16104 (GWverse) and CA18108
(QG-MM) supported by COST (European Cooperation in Science and Technology).
Z. K. was further supported by the János Bolyai Research Scholarship of the
Hungarian Academy of Sciences and by the ÚNKP-20-5-- New National Excellence
Program of the Ministry for Innovation and Technology through its National
Research, Development and Innovation Fund.

\end{document}